\documentclass[preprint,5p]{elsarticle}

\usepackage{lineno}

\journal{Astronomy and Computing}

\usepackage{natbib}
\biboptions{authoryear}
\usepackage{amsmath,mathtools}
\usepackage{amssymb}
\usepackage{txfonts}
\usepackage{here}
\usepackage{xspace}
\usepackage[usenames,dvipsnames]{xcolor}
\newcommand{\Rsun}{\ensuremath{\,\mathrm{R_\odot}}\xspace}
\newcommand{\Msun}{\ensuremath{\,\mathrm{M_\odot}}\xspace}

\definecolor{gray}{HTML}{9E9E9E}

\definecolor{deeppurple}{HTML}{5E35B1}
\usepackage[breaklinks=true,colorlinks,linkcolor=deeppurple,citecolor=blue]{hyperref} 
\usepackage{etoolbox}
\usepackage[]{microtype}
\usepackage{caption}
\usepackage{subcaption}


\begin{document}

\begin{frontmatter}

\title{TULIPS: a Tool for Understanding the Lives, Interiors, and Physics of Stars}

\author{E. Laplace\fnref{myfootnote}}
\address{Anton Pannekoek Institute of Astronomy and GRAPPA, Science Park 904, University of Amsterdam, 1090 GE Amsterdam,\\ The Netherlands.}

\fntext[myfootnote]{\textit{Present address:} Heidelberger Institut f\"ur Theoretische Studien, Schloss-Wolfsbrunnenweg 35, 69118 Heidelberg, Germany}
\ead{e.c.laplace@uva.nl}

\begin{abstract}

Understanding the lives and interior structures of stellar objects is a fundamental objective of astrophysics. Research in this domain often relies on the visualization of astrophysical data, for instance the results of theoretical simulations. However, the diagrams commonly employed to this effect are usually static, complex, and can sometimes be non-intuitive or even counter-intuitive to newcomers in the field. To address some of these issues, this paper introduce \texttt{TULIPS}, a python package that generates novel diagrams and animations of the structure and evolution of stellar objects. \texttt{TULIPS} visualizes the output of one-dimensional physical simulations and is currently optimized for the MESA stellar evolution code. Utilizing the inherent spherical symmetry of such simulations, \texttt{TULIPS} represents the physical properties of stellar objects as the attributes of circles. This enables an intuitive representation of the evolution, energy generation and loss processes, composition, and interior properties of stellar objects, while retaining quantitative information. Users can interact with the output videos and diagrams. The capabilities of \texttt{TULIPS} are showcased by example applications that include a Sun-like star, a massive star, a low-metallicity star, and an accreting white dwarf. Diagrams generated with \texttt{TULIPS} are compared to the Hertzsprung-Russell diagram and to the Kippenhahn diagram, and their advantages and challenges are discussed. \texttt{TULIPS} is open source and free. Aside from being a research tool, it can be used for preparing teaching and public outreach material.
\end{abstract}


\end{frontmatter}

\section{Introduction}
Visualizing observational data and theoretical simulations of the stellar systems that populate the Universe - stars, planets, galaxies, or interstellar clouds - is an essential part of astronomical research \citep{hassan_scientific_2011}. Graphs, maps, charts, diagrams, and sketches are used for the purposes of understanding and explaining concepts, analyzing data, and conveying findings both to peers and to the wider public.

Astronomy has a long history of employing such visualization, \citep[see, e.g.,][]{funkhouser_note_1936}, from abstract to cinematic \citep{aleo_clustering-informed_2020}. 

The field of stellar astrophysics seeks to understand the lifecycle of stars, from their formation to their death. Properties of stars are derived from the analysis of observational data using, e.g., models of their atmospheres or dynamical motion. Such properties in turn serve as constraints for models that describe the dynamical and temporal evolution of single and multiple systems, such as binary star systems, or star clusters. These analyses and simulations rely heavily on data visualization. 
Some visualizations in particular have brought fundamental insights into this field and are now widespread and standard. The Hertzsprung-Russell diagram \citep[HRD,][]{maury_spectra_1897,hertzsprung_uber_1909,russell_relations_1914} revealed that stars follow specific patterns of color and magnitude. Extensive theoretical work and analyses from complementary observational data unveiled the connection between these patterns and the evolutionary stage and mass of stars. Ever since, the HRD has been ubiquitous in stellar astrophysics as a tool to describe and understand the life cycle of stars. Other examples include visualizations of the evolution of the interior energy generation and mixing structure of stars, known as Kippenhahn diagrams (\citealp{hofmeister_sternentwicklung_1964}, see also \citealp{hayashi_evolution_1962}), or visualizations of evolutionary sequences of interacting binary stars \citep{van_den_heuvel_centaurus_1972}.

However, although widespread and standard, many of these visualizations are abstract, static, non-intuitive or even counter-intuitive for those who are new to the field, or suffer from a high density of information.

It is well established in the field of data visualization that the complexity of information displayed directly affects how easily this information can be retained \citep{lusk_effect_1979,evergreen_design_2013} and its general appeal \citep{harrison_infographic_2015}. Evidence suggests that visualizations that are perceived as more appealing convey the information they display more easily \citep{kurosu_apparent_1995,tractinsky_aesthetics_1997}, attract a more diverse audience \citep{korkmaz_primary_2009,harrison_infographic_2015}, and capture the audience's attention for longer \citep{cawthon_effect_2007}. There is supporting evidence that visualizations using real-world objects can convey information more effectively to a diverse audience \citep{lewis_culture_2006}. Dynamic visualizations can make information more understandable and evocative \citep{valkanova_reveal-it_2013} and interactive elements are generally better at capturing and retaining the audience's interest than static ones \citep{newell_picture_2016}. For example, the "Star in a box" education tool \citep{stuart_starinabox_2016} visualizes the life cycle of stars dynamically in a web interface.

This paper introduces the Tool for Understanding the Lives, Interiors and Physics of Stars (\texttt{TULIPS}). It is a python software package that visualizes the evolution and structure of stellar objects based on one-dimensional stellar evolution calculations.
This software addresses some of the important issues of current visualizations employed in stellar astrophysics by introducing a novel approach to visualizing the evolution of stars. \texttt{TULIPS} represents stellar objects as circles, closer to their real-world shape. It visualizes physical properties by employing colors and shadings. In addition, TULIPS enables an interactive visualization of the temporal evolution of these properties by means of animations that can be saved in standard video formats.

Latest state-of-the-art calculations for stellar astrophysics are now wide-spread thanks to the rise of open-source and community-maintained codes, such as MESA \citep{paxton_modules_2011,paxton_modules_2013,paxton_modules_2015,paxton_modules_2018,paxton_modules_2019}. \texttt{TULIPS} visualizes the results of such one-dimensional calculations and is currently optimized for MESA.

The development of this tool began as an effort to better communicate the evolution of stars to students of stellar evolution classes and has evolved into a research tool employed both for analyzing the results of stellar evolution calculations and for communicating these findings in a more intuitive way \citep{laplace_different_2021}. 

\texttt{TULIPS} is open source\footnote{The code can be found at \href{https://bitbucket.org/elaplace/tulips/}{https://bitbucket.org/elaplace/tulips/} and is also listed in the Python Package Index \href{https://pypi.org/project/astro-tulips/}{https://pypi.org/project/astro-tulips/}.} and free for anyone to use as a research, outreach, or education tool (under a GNU general public license, v.3). It is intended to make the structure and evolution of stellar objects more accessible and understandable to a broader community, and as a way to better display and convey the beauty of the physics that governs the lives and interiors of stars. In the spirit of open science, users are encouraged to contribute by sharing the animations and plots they produce with \texttt{TULIPS} with others, and by testing, reporting of bugs, and extending of the capabilities. Contributions can be made by submitting a pull request on the bitbucket repository.

This article is structured as follows. First, the TULIPS software and the underlying concept behind it are described in Section \ref{sec:tulips}. Section \ref{sec:diagrams} and Section \ref{sec:animations} present an overview of the different types of TULIPS diagrams and their animations, respectively. Section \ref{sec:comparison} compares TULIPS to classic diagrams and \ref{sec:disc_conclusion} contains a discussion and conclusion.

\section{\texttt{TULIPS}}
\label{sec:tulips}
\subsection{Basic concept: spherical symmetry}
Assuming that stars are spherically symmetric is at the heart of the classical theory of stellar astrophysics \citep[e.g.,][]{Eddington1926}. This is a good approximation for isolated stellar objects that are mainly subject to self-gravity. Effects of rotation, strong magnetic fields, and binary interactions can lead to deviations from the spherical shape, but for the majority of stellar physics problems, this assumption holds. Furthermore, while these deviations primarily affect the outer layers, spherical symmetry is still a fair approximation for the deeper interior layers of stars where nuclear burning, which drives the life cycle of stars, takes place. 
A powerful consequence of assuming spherical symmetry is that the mathematical description of the structure and evolution of stellar objects is greatly simplified. The entire problem can be captured by a small set of non-linear partial differential equations. The vast majority of stellar evolution calculations consist in solving these one-dimensional equations with various numerical methods and physical assumptions \footnote{Multi-dimensional calculations of stellar evolution also exist \citep[recent examples include e.g.,][]{arnett_turbulent_2009,fields_development_2020,yadav_large-scale_2020}. However, due to the extreme spatial and time scales involved, only short portions of stellar lives can be computed.}.

The basic concept behind \texttt{TULIPS} is to make use of the intrinsic spherical symmetry to represent any physical (one-dimensional) property of a star as the radius, surface area or color of a (two-dimensional) circle. 
Physical properties of stars are generally expressed as a function of their position $r$ (known as Eulerian coordinate), where $r$ varies from 0 at the center of the star to the total radius R at the surface. They can also be expressed as a function of the mass $m$ of a small shell inside the star (known as Lagrangian coordinate), where $m$ varies from 0 at the center to the total mass M at the surface. \texttt{TULIPS} can visualize the physical properties of stars in either form by representing stellar models as circles whose radius represents one or the other coordinate.

\subsection{The \texttt{TULIPS} software}
The first version of the \texttt{TULIPS} python software is written with functional design. To create diagrams or animations, users call specific functions. \texttt{TULIPS} contains a set of core functions that create visualizations of stellar properties. By default, these functions create a visualization at a fixed point in time (see Section \ref{sec:diagrams}). By changing the time argument, the same functions can be used to create animations from a start point to an end point (see Section \ref{sec:animations}). \texttt{TULIPS} relies on \texttt{matplotlib} functionalities to create plots and animations that can be saved in standard image or video formats. The visual aspect of \texttt{TULIPS} diagrams and animations can be customized easily. Basic and advanced use of the \texttt{TULIPS} is described in detailed in the documentation and tutorials\footnote{The documentation and tutorials can be found at \href{https://astro-tulips.readthedocs.io/en/latest/}{https://astro-tulips.readthedocs.io/en/latest/}}. TULIPS builds upon several existing open-source python packages, including \texttt{mesaPlot} \citep{farmer_mesaPlot_2019}, \texttt{matplotlib} \citep{hunter_matplotlib_2007}, \texttt{numpy} \citep{van_der_walt_numpy_2011}, \texttt{ColorPy}\footnote{\copyright Mark Kness, \href{http://markkness.net/colorpy/ColorPy.html}{http://markkness.net/colorpy/ColorPy.html}}, \texttt{astropy} \citep{astropy:2013, astropy:2018}, \texttt{CMasher} \citep{van_der_velden_cmasher_2020}, and \texttt{ipython/jupyter} \citep{perez_ipython_2007}.

\subsection{Input physical simulations}
\texttt{TULIPS} itself does not solve the stellar structure equations. Instead, it uses the solutions generated by existing simulations as input to generate diagrams and animations. Although \texttt{TULIPS} can be adapted for use with any one-dimensional simulation, in principle, it has been optimized for use with the open-source Modules for Experiments in Stellar Astrophysics \citep[MESA, version 15140;][]{paxton_modules_2011,paxton_modules_2013,paxton_modules_2015,paxton_modules_2018,paxton_modules_2019}. The example stellar evolution models shown in this work are based on default stellar models from the MESA test suite and on models from \citet{laplace_different_2021}. MESA inlists and data products will be made available upon publication\footnote{The data is available at \href{https://doi.org/10.5281/zenodo.5032250}{https://doi.org/10.5281/zenodo.5032250}}.

For reading output files from \texttt{MESA} simulations, \texttt{TULIPS} uses the open-source python package \texttt{mesaPlot} \citep{farmer_mesaPlot_2019}. With this package, the entire information contained in output files from a \texttt{MESA} computation can be stored in a single python object that allows easy access to all physical quantities contained in the output files. All core \texttt{TULIPS} functions that create visualizations require this object as an argument. Depending on the information they visualize, \texttt{TULIPS} functions necessitate different types of MESA output files. Two types of output files exist: MESA \textit{history} files include the evolution of one-dimensional physical properties as a function of time, while MESA \textit{profile} files are snapshots of the interior properties of stellar objects as a function of their mass coordinate at one particular moment in time. The names of functions that require the latter output type contain the suffix \texttt{\_profile}.

\section{\texttt{TULIPS} diagrams: overview}
\label{sec:diagrams}


\begin{figure*}[!ht] 
	\begin{subfigure}[t]{0.5\textwidth}
		\includegraphics[width=\textwidth]{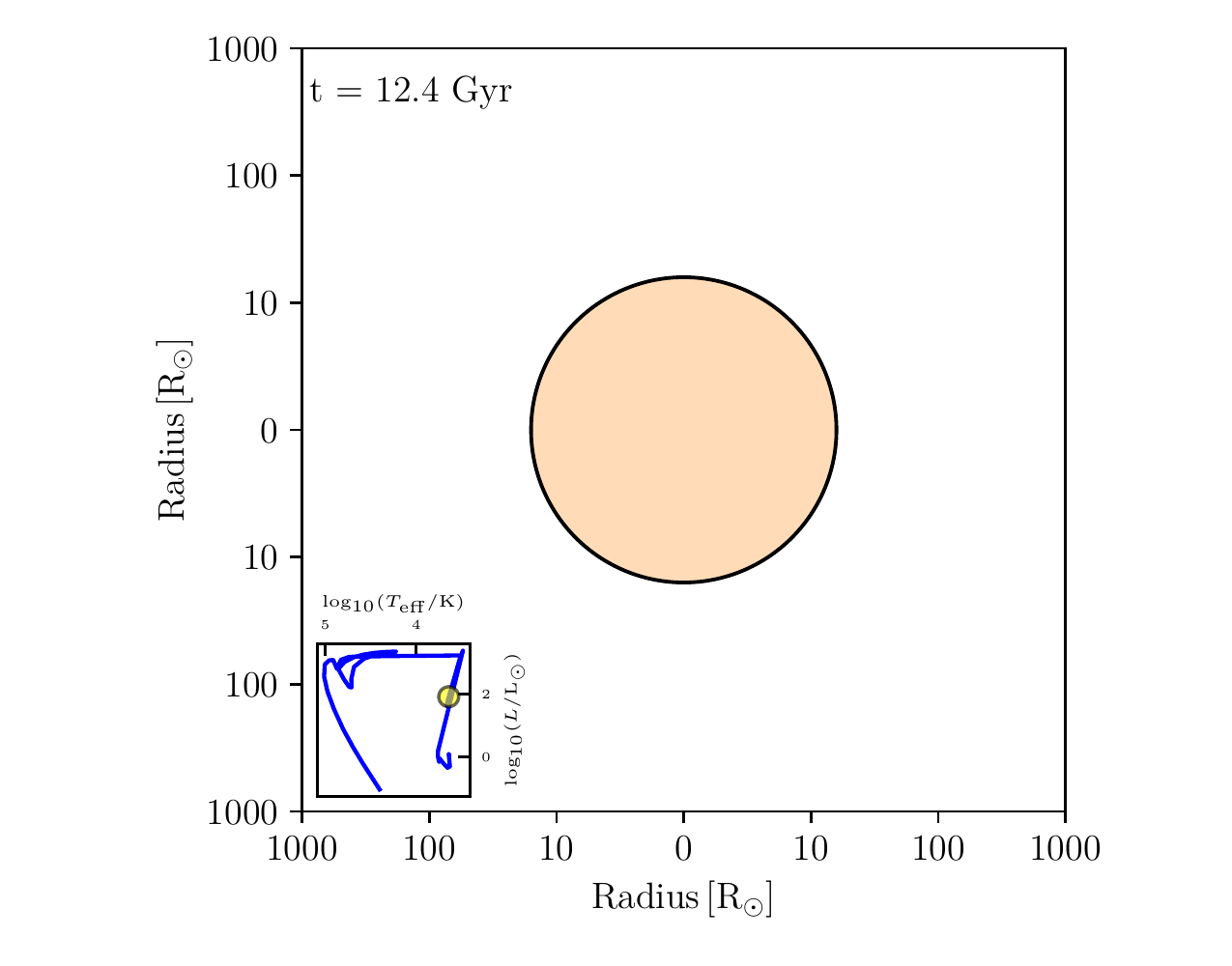}
		\caption{\texttt{perceived\_color} diagram}
	\end{subfigure}
	\begin{subfigure}[t]{0.5\textwidth}
		\includegraphics[width=\textwidth]{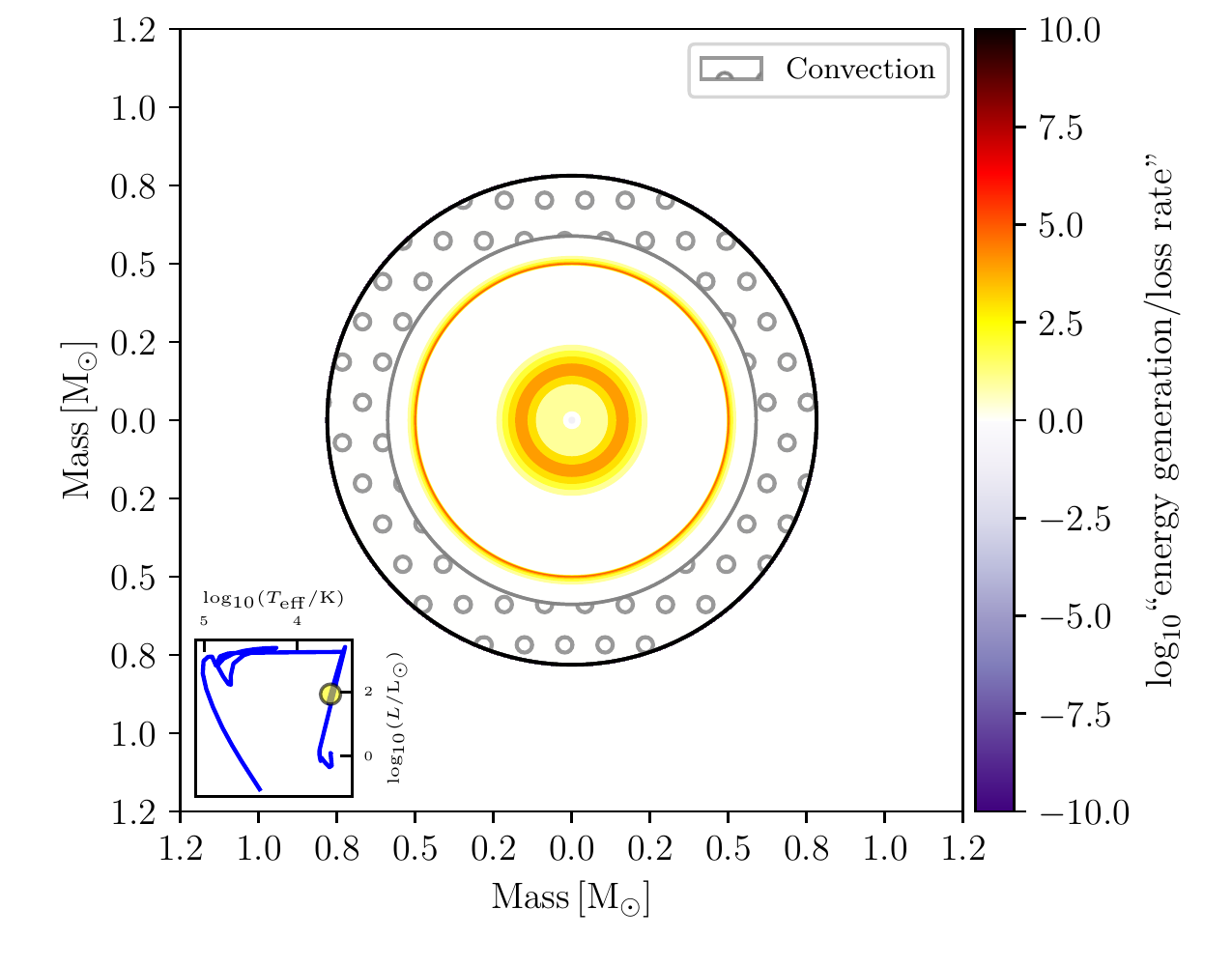}
		\caption{\texttt{energy\_and\_mixing} diagram}
	\end{subfigure}
	\begin{subfigure}[b]{0.5\textwidth}
		\includegraphics[width=\textwidth]{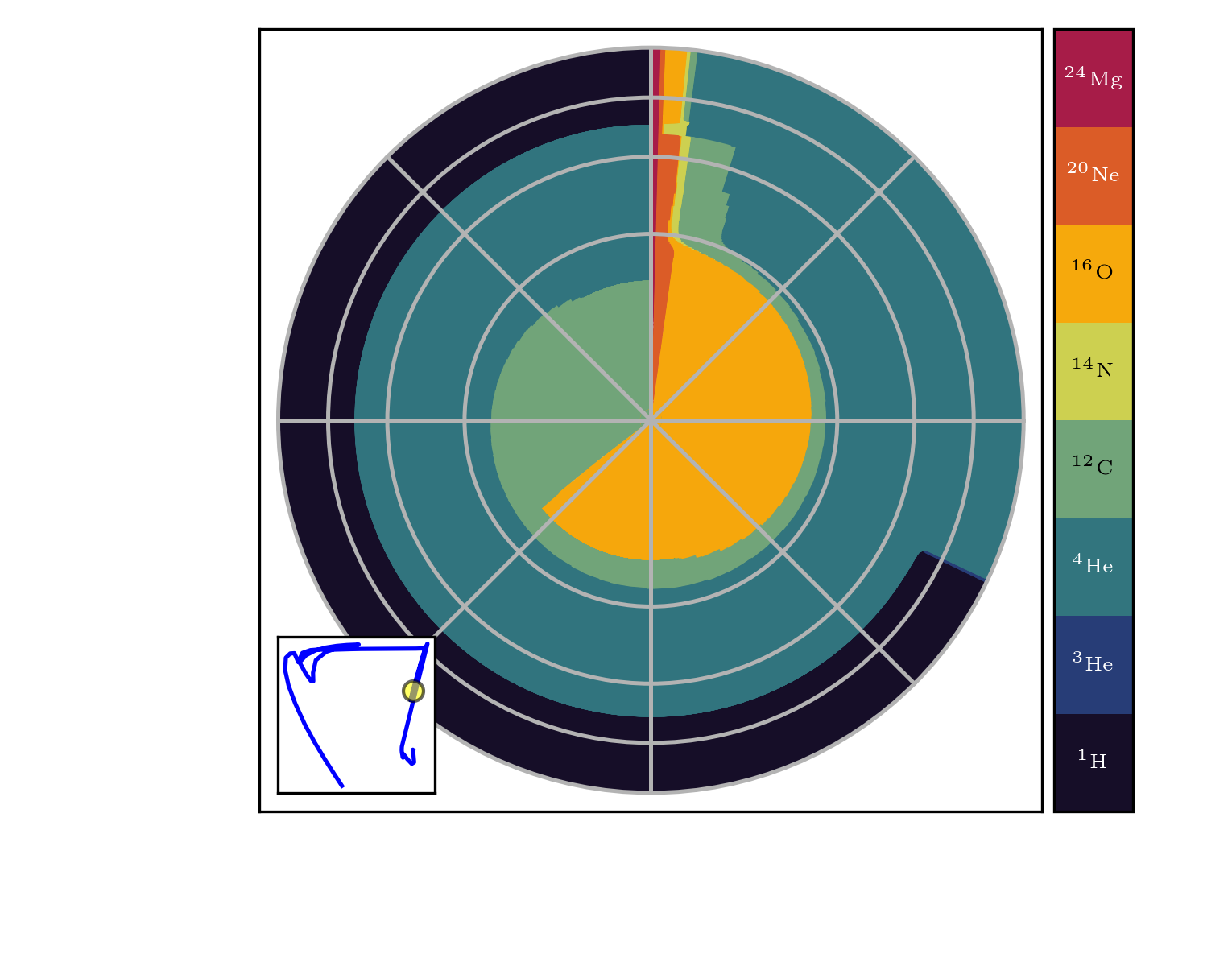}
		\caption{\texttt{chemical\_profile} diagram}
	\end{subfigure}
	\begin{subfigure}[b]{0.5\textwidth}
		\includegraphics[width=\textwidth]{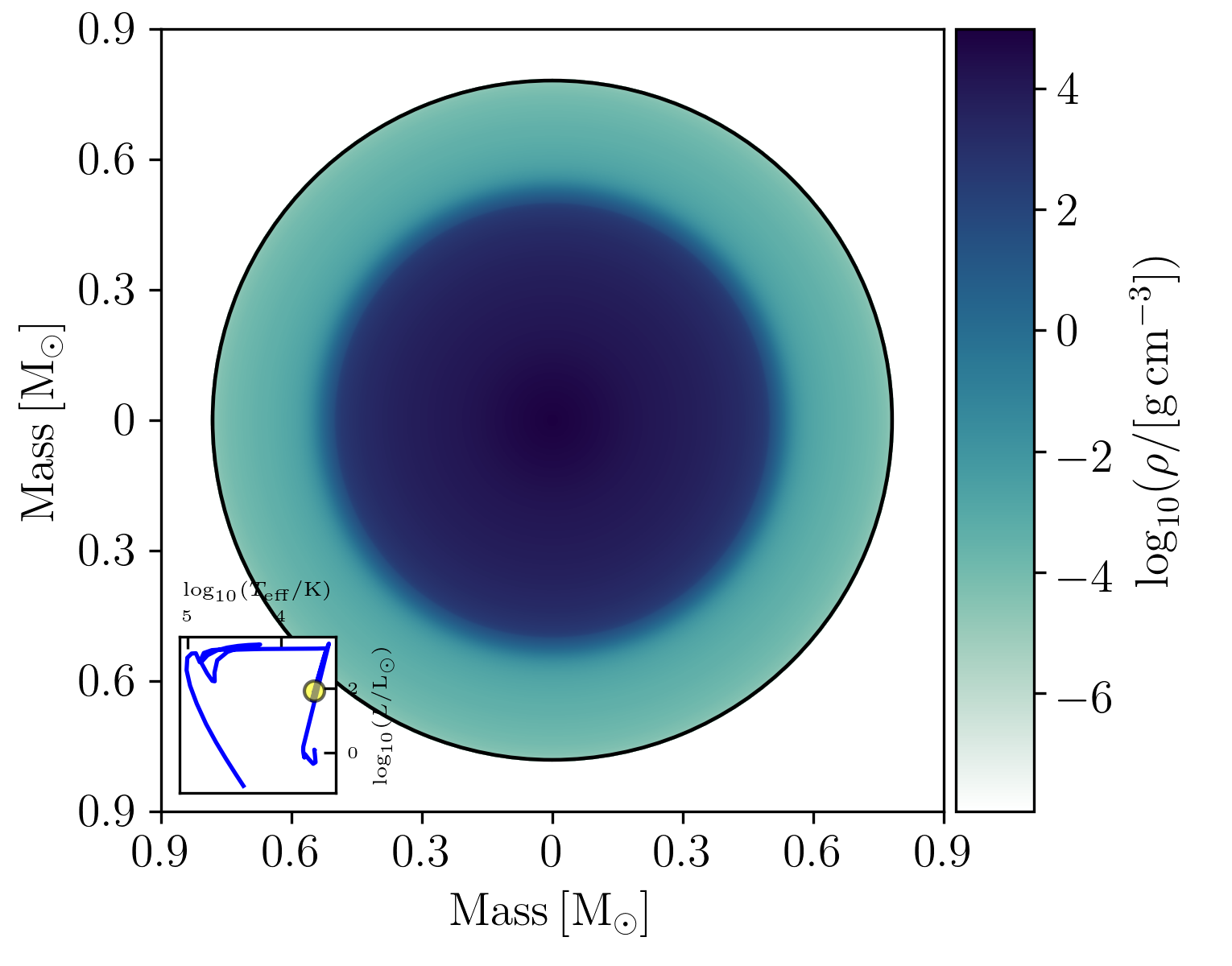}
		\caption{\texttt{property\_profile} diagram}
	\end{subfigure}
	\caption{Overview of the four basic \texttt{TULIPS} diagrams for a Sun-like star at the end of core helium burning. Inset diagrams at the bottom left of each panel show the location of the star on the HRD. \textbf{a.} The star has a radius of about 12 \Rsun and appears orange, as expected for a red clump star. \textbf{b.} The star contains a hydrogen-burning shell and a helium-burning core. In the very center, energy losses can be observed. \textbf{c.} From the center outwards, the star is composed of carbon and oxygen, followed by a layer that predominantly contains helium, and then by an extended envelope mainly composed of hydrogen and helium. Here, the radial direction is proportional to the square root of the mass of the star. Overlayed circles give the location of particular mass fractions of the star. From the center, moving outward, these are 0.25, 0.5, 0.75, and 1 times total mass of the star \textbf{d.} The density throughout the stellar interior. The star contains a dense stellar core surrounded by a lower-density region, and has a low-density envelope.} 
	\label{fig:sun_overview}
\end{figure*}

Fig. \ref{fig:sun_overview} showcases the four basic diagrams that \texttt{TULIPS} produces for the default MESA stellar model of a Sun-like star (\texttt{1M\_pre\_ms\_to\_wd} model in the MESA test suite). The diagrams are all generated at the end of core helium burning (when the central mass fraction of helium becomes lower than $10^{-4}$).
By calling the corresponding function, the following TULIPS diagrams are shown in the Figure, from left to right and top to bottom:
\begin{enumerate}[a.]
	\item \texttt{perceived\_color}: a diagram that shows a physical property of the entire star, for example its total radius and the (approximate) color of a star as perceived by the human eye.
	\item \texttt{energy\_and\_mixing}: a diagram that shows the energy generation and losses, and the mixing processes in stellar objects.
	\item \texttt{chemical\_profile}: a diagram that represent the interior composition profile of a star.
	\item \texttt{property\_profile}: a diagram that represent an internal physical property, for example the density as a function of the mass coordinate.

\end{enumerate}
In addition, \texttt{TULIPS} allows comparisons with typical diagnostic diagrams used in stellar astrophysics, such as the HRD, by including these as insets in the corner of a \texttt{TULIPS} diagram.

\subsection{Diagram a: a star's radius and its color as perceived by the human eye}
\label{sec:diagrams:radius_color}
It can be challenging to apprehend the observable properties of a stellar object at a particular evolutionary stage or at a particular location on the HR diagram. \texttt{TULIPS}' \texttt{perceived\_color} diagrams help address this by representing basic properties of a stellar object as the radius and color of a circle, where the color corresponds to the (approximate) color of a star as perceived by the human eye\footnote{In reality, at close distance stars are so luminous that most colors would be saturated and appear as white to the human eye.}. In the example shown in the panel a of Fig. \ref{fig:sun_overview}, the diagram visualizes how a Sun-like star that has just completed central helium burning and has an effective temperature of $4395\,\mathrm{K}$ may be perceived by the human eye: the star is represented as an orange circle with a radius of 12 \Rsun, which correspond to a red clump star. The location of the star on the HRD at this evolutionary stage is shown in an inset located in the bottom left corner of the diagram.

With these diagrams, \texttt{TULIPS} can represent not only the radius, but any one-dimensional physical property of  a star (for example its mass or density) as the radius of a circle. To compute the (approximate) human-perceived color of the stellar object, \texttt{TULIPS} makes use of the \texttt{ColorPy} python package. With this tool, the intensity spectrum of a star is approximated as the blackbody spectrum expected for a given effective temperature. This spectrum is converted to the approximate range of colors a human eye can perceive by using the 1931 color matching functions of the Commission Internationale de l'\'Eclairage \citep{smith_cie_1931}. This process results in a color for each effective temperature.
Animations of this diagram visualize how the stellar property shown (e.g., the radius) varies as a function of time and can help better appreciate the scales involved (see Section \ref{sec:animations:HRD}).

\subsection{Diagram b: energy generation/losses and mixing in the stellar interior}
\label{sec:diagrams:energy_mixing}
The \texttt{energy\_and\_mixing} diagram helps to visualize the interior properties of stars at a certain evolutionary stage, at a particular location on the HR diagram, or when a certain condition is met (for example, when the central temperature of the star reaches a certain value). Energy generation and losses are indicated with colors \citep[as computed by \texttt{mesaPlot}, see also][]{farmer_carbon_2015}. Optionally, various types of mixing that occur in the stellar interior can be indicated with hatched regions. The example shown in the panel b of Fig.~\ref{fig:sun_overview} presents the interior burning and mixing processes of a Sun-like star at the end of core helium burning. The radius of the outermost circle gives the total mass of the star, which is just under a solar mass at this moment due to previous mass loss during the giant phase. Two sets of yellow and orange rings indicate shell burning: helium shell burning near the center and a weaker hydrogen-burning shell further out. The purple circle in the very center represents energy losses in the innermost region due to neutrino emission. Grey rings containing small gray circles indicate that convection is occurring in the outermost layers of the star. For simplicity, the energy/loss rate $\epsilon$ is not written out in the colorbar label. This quantity is computed as follows:
$\epsilon = \textrm{sign}(\epsilon_{\mathrm{nuc}} - \epsilon_{\nu})\textrm{log}_{10}(\textrm{max}(1.0,|\epsilon_{\mathrm{nuc}}-\epsilon_{\nu}|)/[\textrm{erg g}^{-1}\textrm{s}^{-1}])$, where $\epsilon_{\mathrm{nuc}}$ is the nuclear burning rate and $\epsilon_{\nu}$ the neutrino energy loss rate. These diagrams contain similar information as Kippenhahn diagrams for a particular moment in time. Animations of \texttt{energy\_and\_mixing} diagrams contain similar information as Kippenhahn diagrams (see also Section \ref{sec:comparison:kipp}). 

\subsection{Diagram c: the star's interior composition}
\label{sec:diagrams:composition}
For many problems in stellar physics, it is important to know the interior composition of a stellar object at a particular moment in time. With \texttt{TULIPS}, this information can be displayed as the property of a circle that contains nested pie charts, as first introduced in \citet{laplace_different_2021}. Each of these pie charts represent the mass fractions of isotopes at a particular coordinate in the stellar interior. Each isotope correspond to a particular color, as shown in the legend. By default, the radial direction is proportional to the square root of the mass coordinate. As a result, the surface area spanned by a certain color is proportional to the total mass of the isotope it represents in the star. Fig. \ref{fig:sun_overview} contains an example composition diagram for a Sun-like star at the end of central helium burning in the panel c. From outside, moving inward, it is composed by a hydrogen-rich envelope that contains mass fractions of 0.7, 0.25, and 0.01 of hydrogen, helium, and heavier elements, respectively, as expected for a Solar-like composition \citep{asplund_chemical_2009}. Below the envelope, at 75\% of the total mass of the star, the model contains a helium-rich layer with low fractions of carbon, oxygen, and neon. The mass of the innermost core is divided in 0.33 carbon and 0.66 oxygen, and low fractions of neon and magnesium. Overlayed grey circles and lines help read quantitative information from these composition diagrams. For example, they clarify that the boundary of the innermost core coincides with about a quarter of the total mass of the star.

To create these diagrams, \texttt{TULIPS} automatically identifies all isotopes present in the stellar interior\footnote{A list of isotopes of interest can also be specified by the user.} and assigns them a color based on a custom colormap. For efficiency, the stellar interior is then divided (down-sampled) into a number of concentric rings (by default, 300). The code then interpolates the changes in composition at the location of these rings. For each ring, \texttt{TULIPS} constructs a pie chart based on the mass fractions of isotopes within the mass or radius extent of the ring. By default, all pie charts start at a 12 o'clock position. The isotopes are shown counter-clockwise in order of increasing mass number. For a typical stellar object composed primarily of hydrogen, helium, and ``metals" this means the pie chart begins with $^{1}\mathrm{H}$, followed by $^{4}\mathrm{He}$, and ends with an iron-group element, such as $^{56}\mathrm{Fe}$. As a consequence, when animated as a function of time, the growing mass fraction of isotopes in the stellar interior (for example helium during core hydrogen burning) produce a clockwise motion in the composition diagrams (see the example animation described in Section \ref{sec:animations:HRD} in the supplementary material).

\subsection{Diagram d: a physical property throughout the stellar interior}
\label{sec:diagrams:property}
\texttt{TULIPS}' \texttt{property\_profile} function visualizes how a certain physical property, such as the density, changes from the center to the surface of a stellar object. These type of diagrams are commonly referred to as the profile of such a property. An example of a \texttt{TULIPS} diagram representing the density profile of a Sun-like star at the end of core helium burning is shown in the panel d of Fig.~\ref{fig:sun_overview}. The star is divided into multiple rings where each ring corresponds a zone within the stellar model (here, 1000 zones). The radii of all rings are chosen such that their location is proportional to the enclosed mass of the star. The color of each ring corresponds to the density $\rho$ of the star, as specified by the color scale. In this example, the outer layers have a very low density of $\log_{10} (\rho/[\mathrm{g}\,\mathrm{cm}^{-3}]) = -5$, while the core has a higher density of the order of $\log_{10} (\rho/[\mathrm{g}\,\mathrm{cm}^{-3}]) = 3$ and is surrounded by a lower density ring of about $\log_{10} (\rho/[\mathrm{g}\,\mathrm{cm}^{-3}]) = -2$.

\subsection{Combining multiple \texttt{TULIPS} diagrams}
\label{sec:diagrams:combinations}
\begin{figure}[ht!] 
	\includegraphics[]{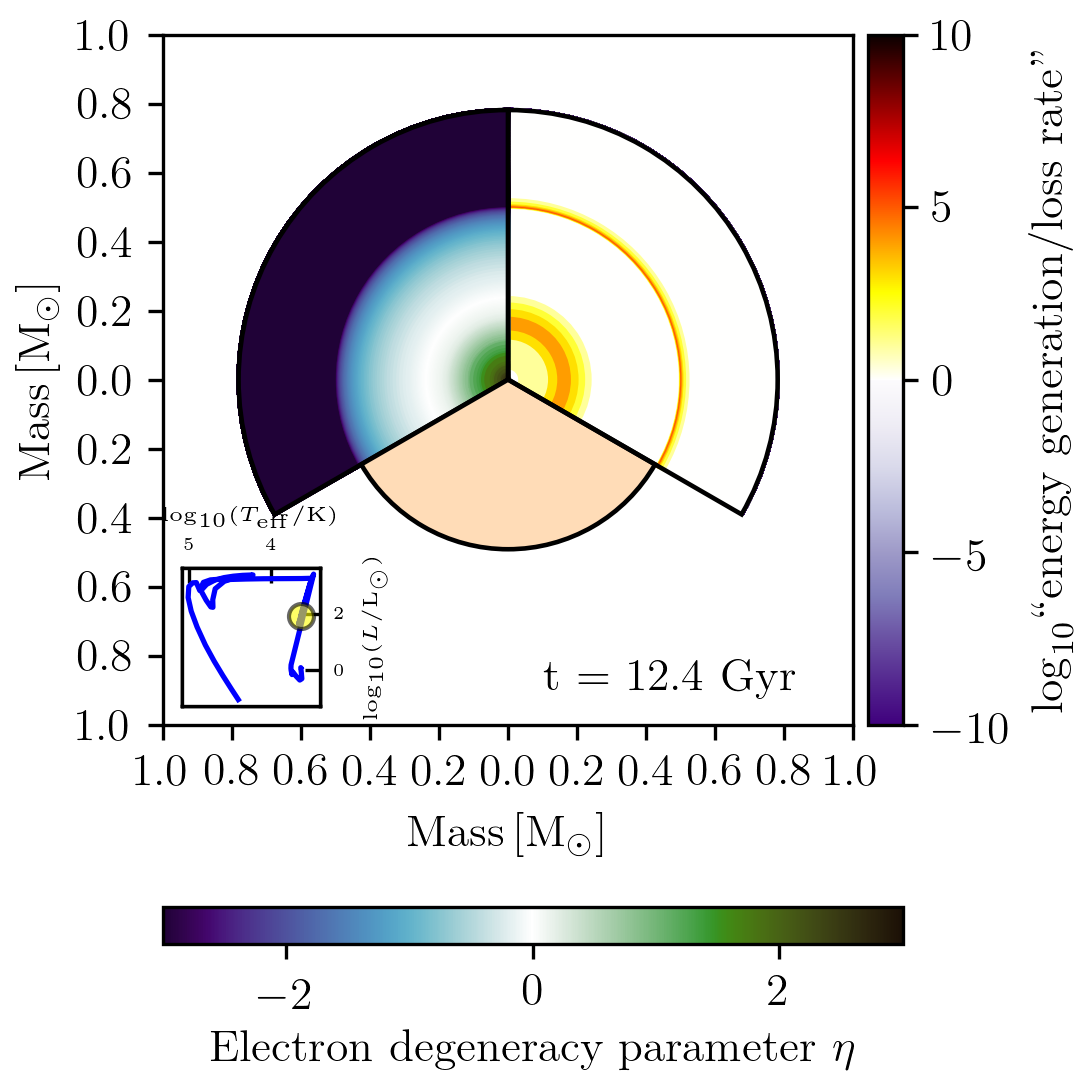}
	\caption{Example combination of TULIPS diagrams in one plot. All diagrams represent properties of the same model of a Sun-like star at the end of core helium burning. From left to right, the dimensionless electron degeneracy parameter $\eta$ \citep[electrons are degenerate when $\eta > 0$,][]{paxton_modules_2011}, the mass of the helium core with the perceived color of the star, and energy generation processes in the stellar interior, are shown.} 
	\label{fig:sun_combined}
\end{figure}

Because the information displayed in diagrams a, b, and d is independent from the angle chosen, these diagrams can be combined in the same figure. This is demonstrated in Fig. \ref{fig:sun_combined}, where information from three different types of TULIPS diagrams for the same Sun-like stellar model at core helium depletion as shown in Fig.\ref{fig:sun_overview} is combined. All diagrams display stellar properties as a function of the mass coordinate. This figure reveals that the outer, hydrogen-burning shell of the star is located just above the edge of the helium-rich core. It also demonstrates that electrons are degenerate in the very center of the star (degeneracy parameter $\eta$ larger than zero), below the helium-burning region. The figure shows that the electron degeneracy varies greatly between the outer envelope and the inner helium-rich core of the star. For each diagram, we specify a different starting and end angle. This feature of \texttt{TULIPS} can also be used to compare of different MESA models at similar evolutionary stages (for an example, see Fig. 13 in \citealp{laplace_different_2021}).

\section{TULIPS animations}
\label{sec:animations}
Animations are the centerpiece of \texttt{TULIPS}' capabilities. Every diagram produced with \texttt{TULIPS} can be animated. This enables an interactive exploration of the properties of stellar objects as a function of time and can help the users gain more insight into the meaning of particular features in classic diagrams, such as the HRD or Kippenhahn diagrams.

\texttt{TULIPS} animations are created with the \texttt{matplotlib} Animation module \citep{hunter_matplotlib_2007}.
\texttt{TULIPS} offers multiple options to adapt the time step and frame rate of the resulting animation (see documentation). By default, the time step follows the MESA \texttt{model\_number} of a simulation, a number that keeps track of models generated in a MESA calculation. In addition, options that follow the evolutionary time are also available. Because of the vastly different time scales involved in the various evolutionary stages of a star (for a typical single massive star, millions of years during the main sequence compared to days for the last burning stages, \citealt{woosley_evolution_2002}), it is often useful to re-scale the evolutionary time such that most evolutionary phases will span a similar duration in the animation. This scaling can be applied in \texttt{TULIPS} animations by using the \texttt{log\_to\_end} time scaling option. An option that follows the age of the star linearly is also available. For TULIPS functions that rely on MESA profile output, it is essential to have a larger number of profiles generated at regular intervals to accurately capture the time evolution (for more details, see the TULIPS documentation).

\subsection{Understanding features on the HRD tracks of stellar objects}
\label{sec:animations:HRD}
\begin{figure*}[ht] 
	\includegraphics[width=0.33\textwidth]{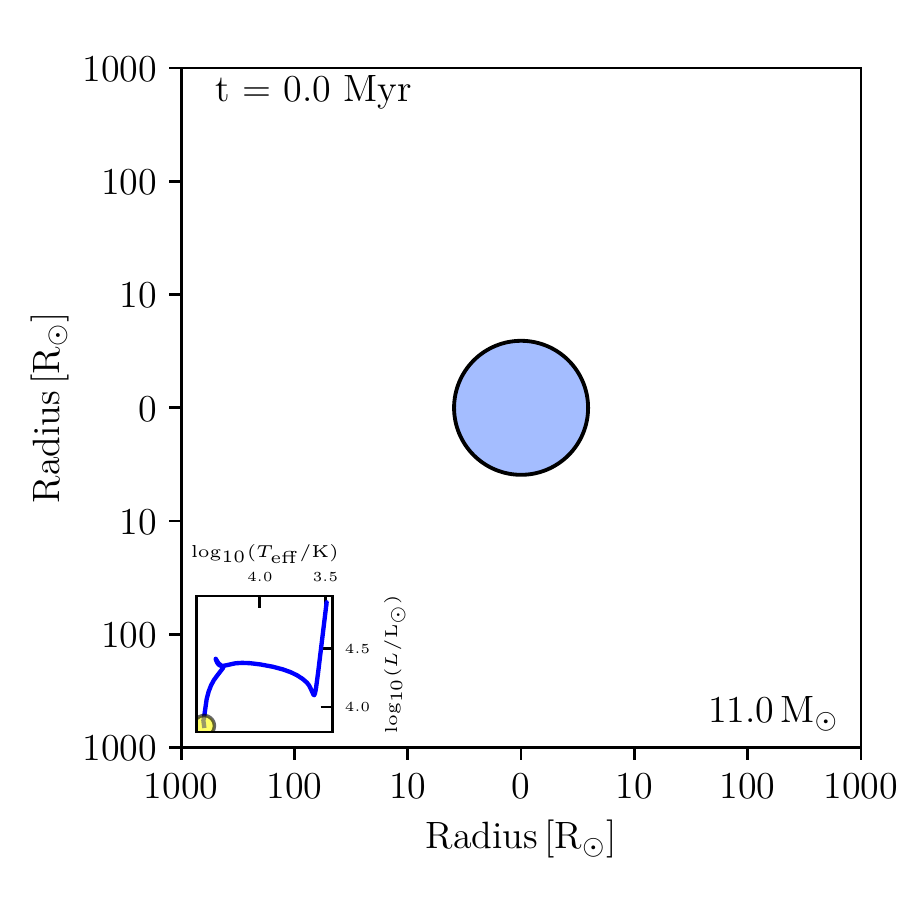}%
	\includegraphics[width=0.33\textwidth]{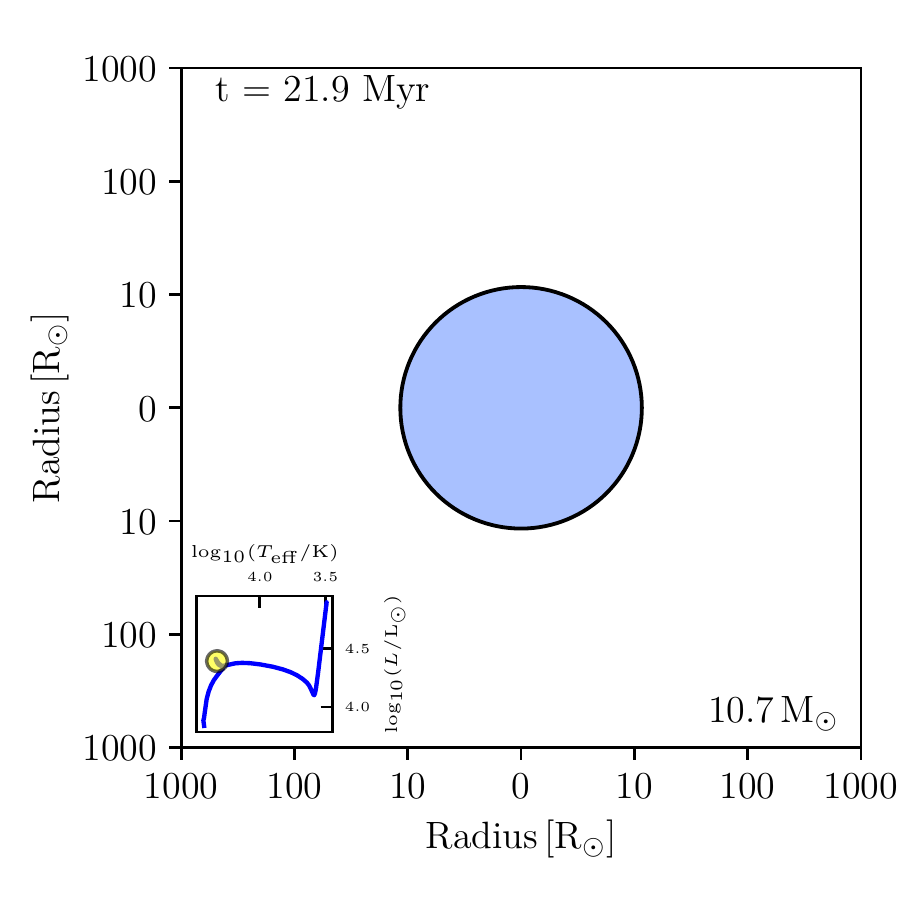}%
	\includegraphics[width=0.33\textwidth]{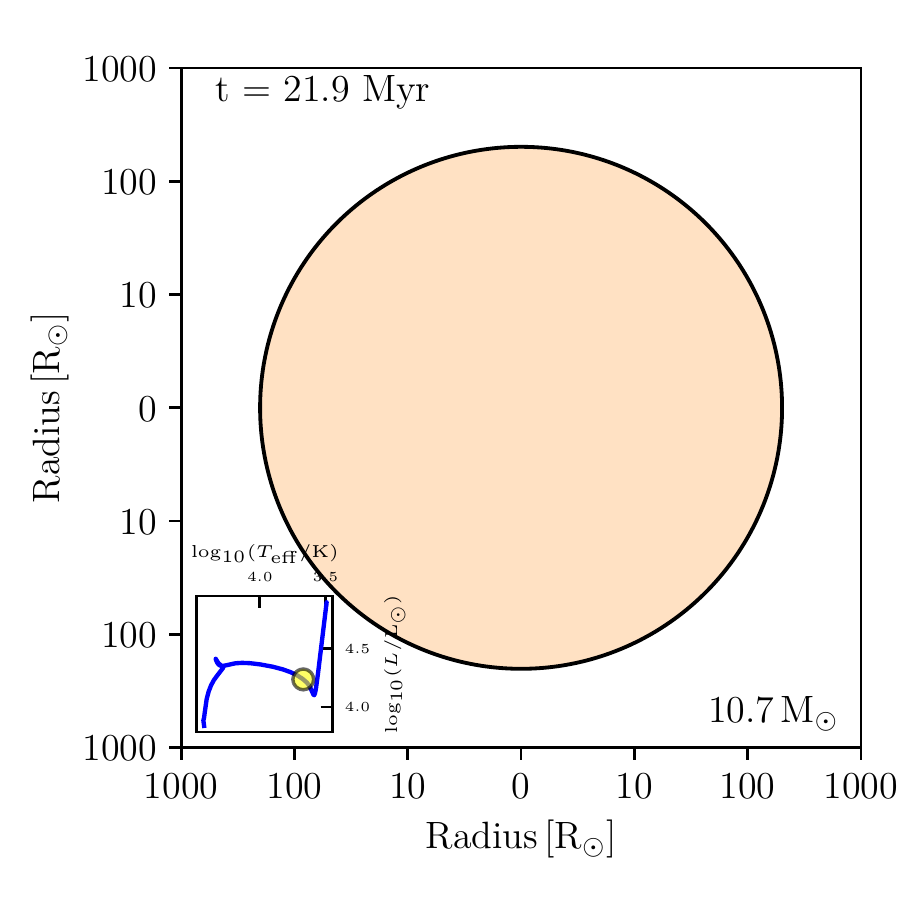}
	\includegraphics[width=0.33\textwidth]{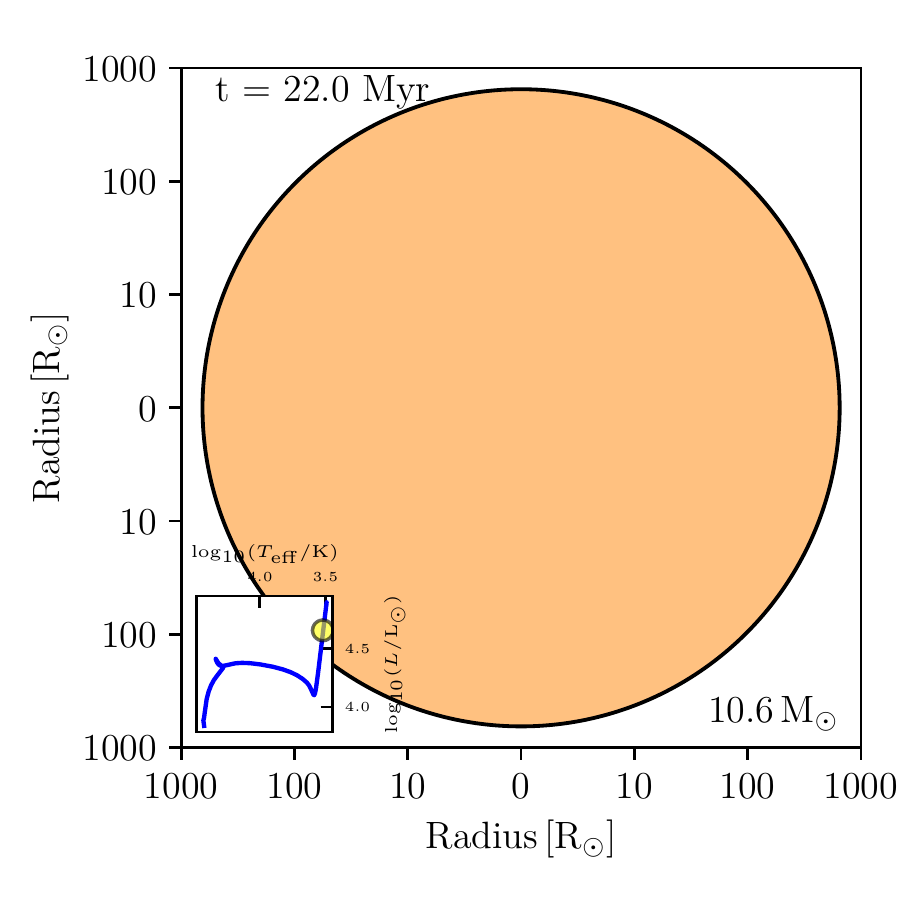}%
	\includegraphics[width=0.33\textwidth]{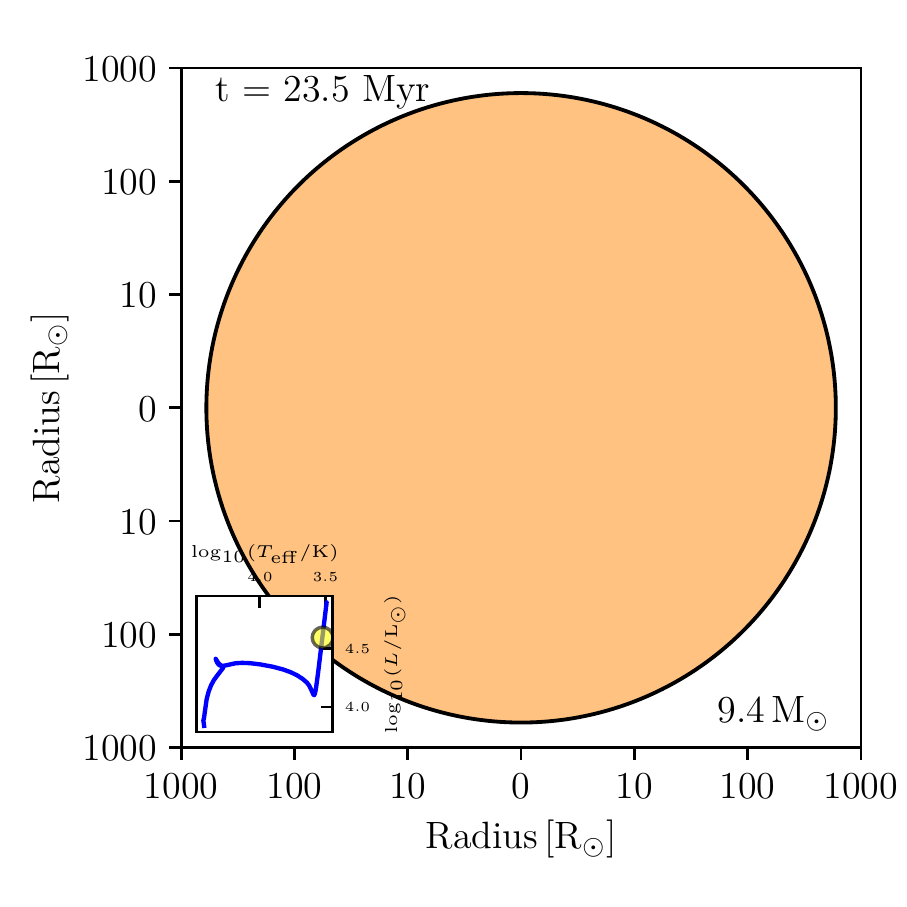}%
	\includegraphics[width=0.33\textwidth]{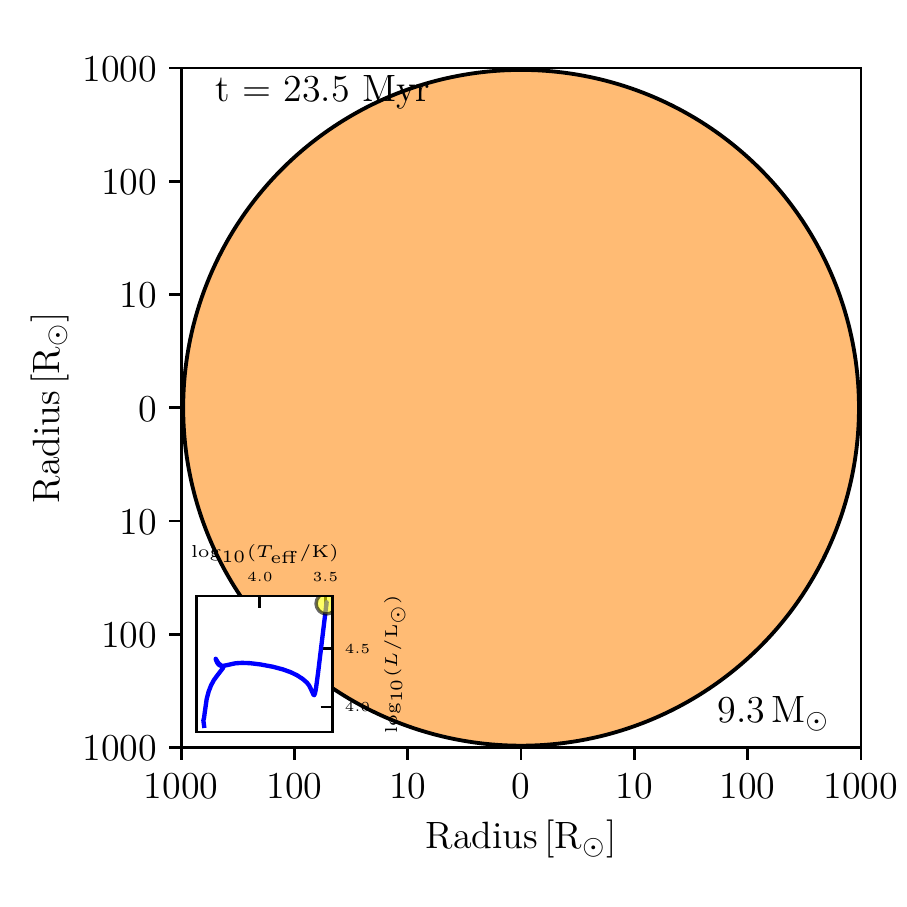}%
	\caption{Snapshots of a \texttt{TULIPS} animation that shows the evolution of a single 11\Msun star at solar metallicity from the onset of core hydrogen burning to the end of core oxygen burning. The diagrams represent the perceived color and radius evolution of the star. Inset diagrams indicate the evolution on the HRD.}
	\label{fig:hrd_m11}
\end{figure*}

\texttt{TULIPS} animations can help gain a better understanding of particular evolutionary stages in the lives of stellar objects and to develop an intuition for tracks on the HRD. In Fig.~\ref{fig:hrd_m11}, a series of snapshots from an example animation is shown. The animation features the radius and perceived color evolution of an 11\Msun single star at solar metallicity (see Section \ref{sec:diagrams:radius_color}). Inset diagrams in the lower left corners help compare the evolution of this massive star on the HRD. In step 1, the star starts its evolution on the main sequence and experiences a slow radial expansion until hydrogen is exhausted in the core in step 2. In step 3, the star experiences a sudden increase in size of two orders of magnitude after having left the main sequence, and changes in perceived color, becoming a red supergiant. In step 4, the star slowly climbs the red supergiant branch and expands slightly. Between step 4 and step 5, the stellar radius decreases again as core helium burning begins. This radius evolution can easily be missed on the HRD because the evolutionary track on the HRD overlaps with the previous evolution. After this moment, the star expands again slowly before reaching its final location on the HRD, shown in step 6.

\subsection{Animations to develop physical intuition: accretion onto a white-dwarf leading to stable hydrogen burning}
\label{sec:animations:wd}

\begin{figure*}[ht] 
	\includegraphics[width=0.33\textwidth]{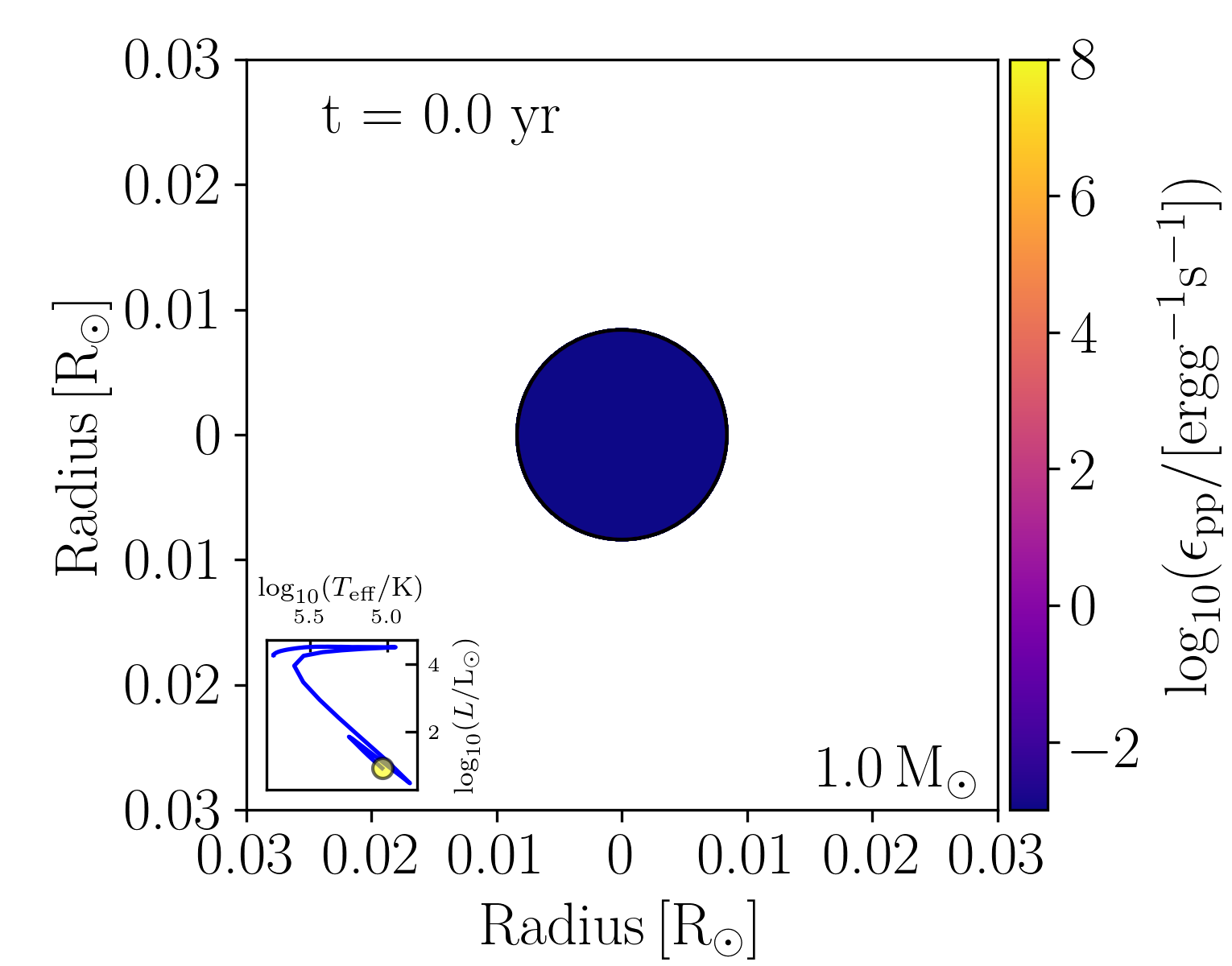}%
	\includegraphics[width=0.33\textwidth]{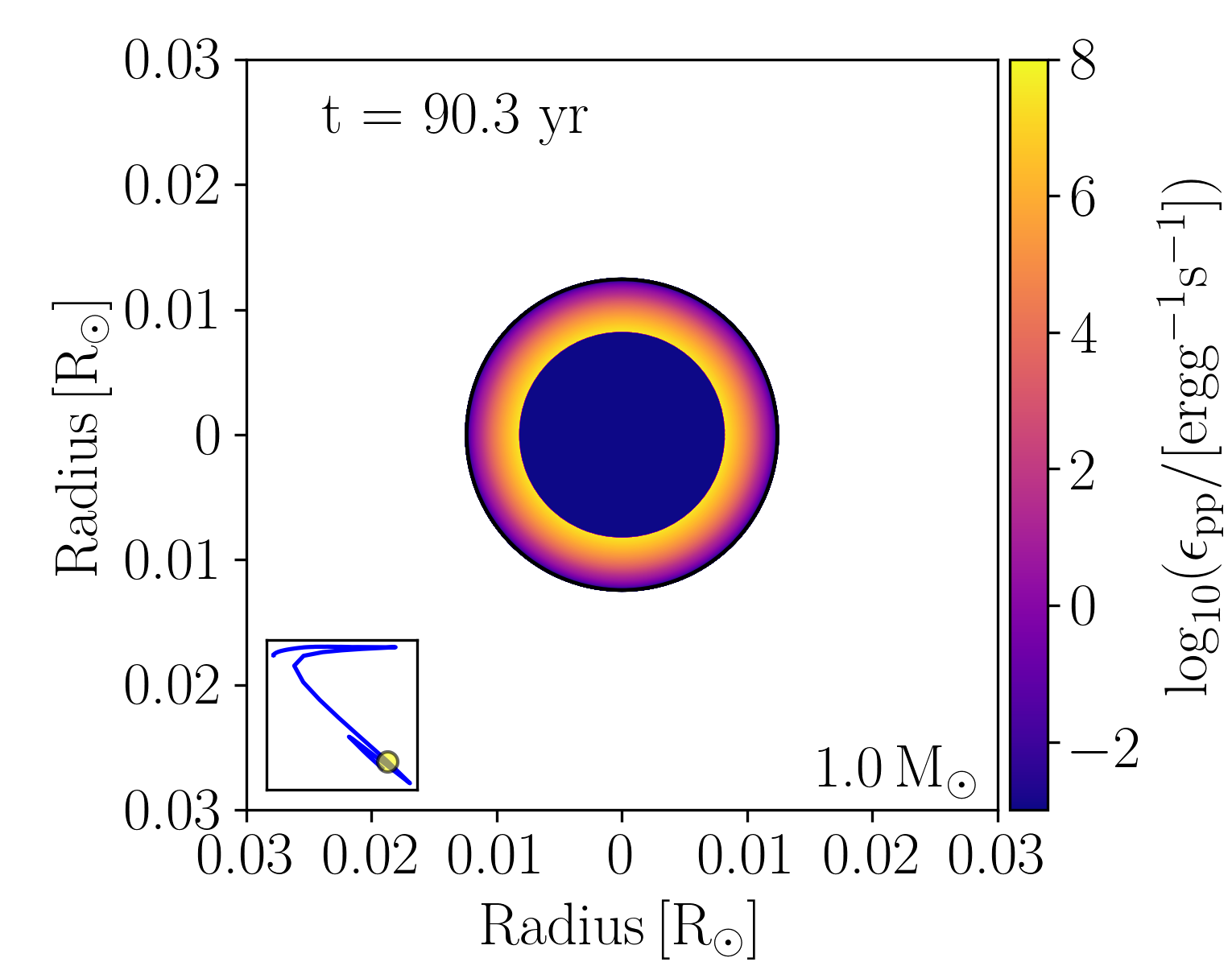}%
	\includegraphics[width=0.33\textwidth]{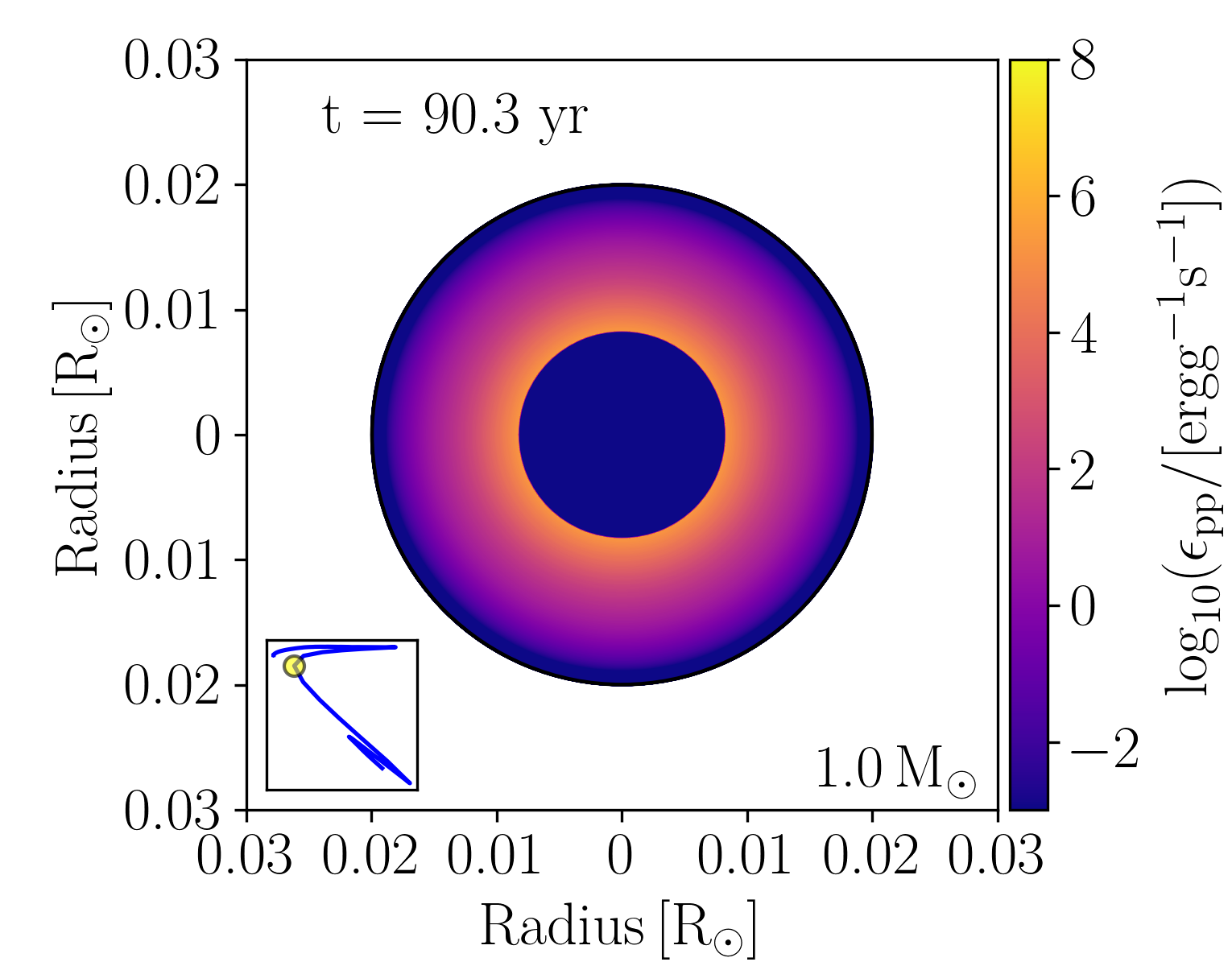}
	\includegraphics[width=0.33\textwidth]{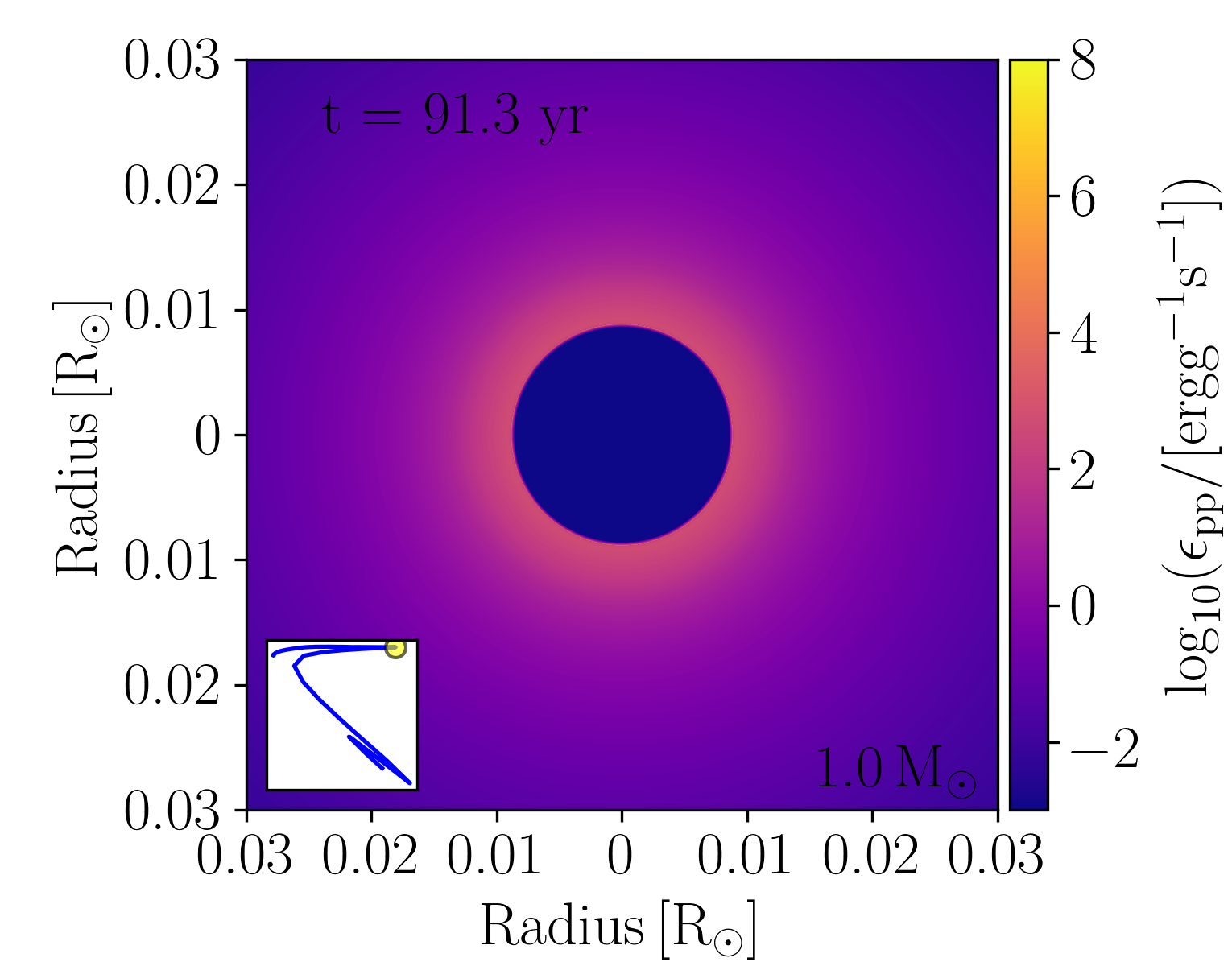}%
	\includegraphics[width=0.33\textwidth]{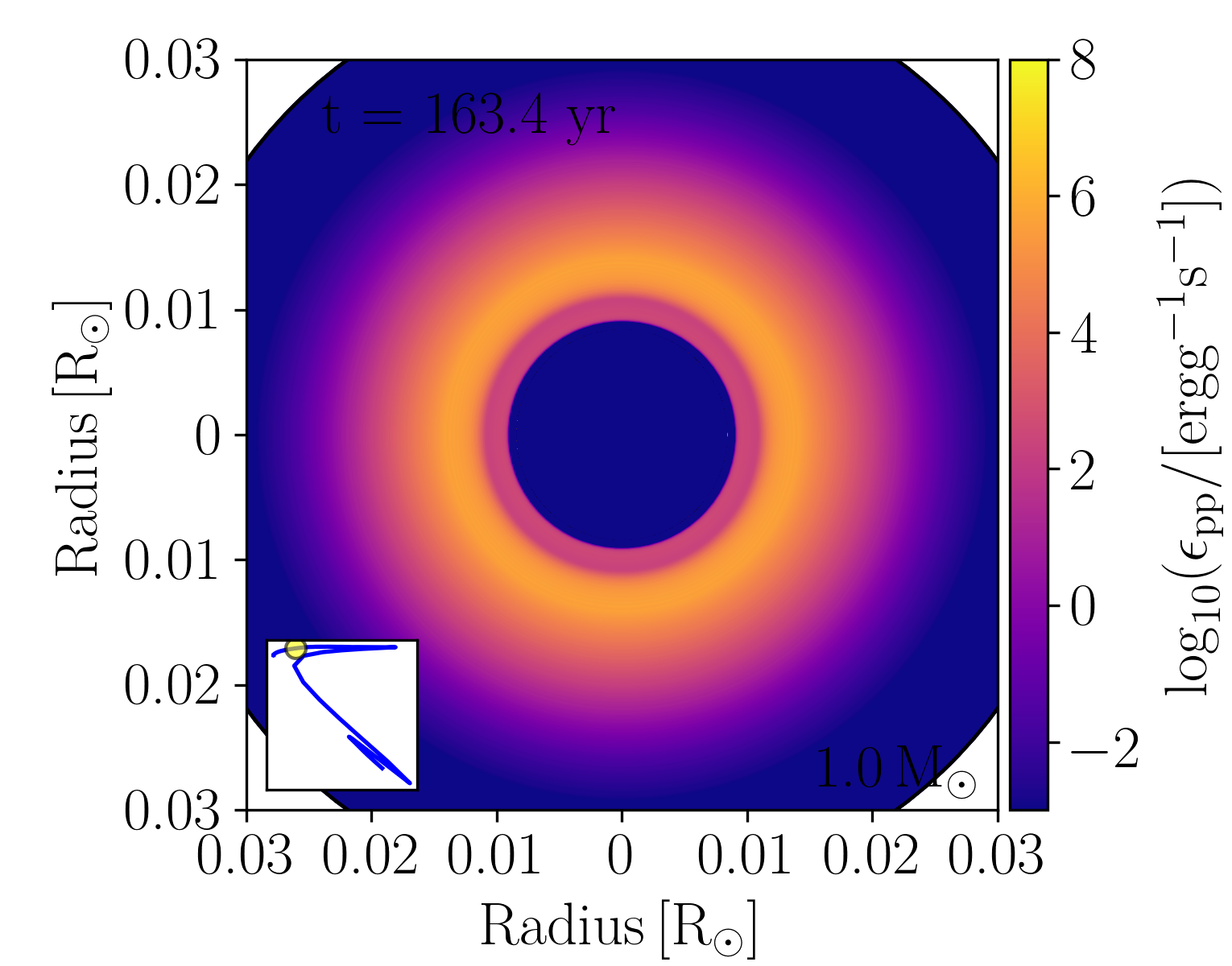}%
	\includegraphics[width=0.33\textwidth]{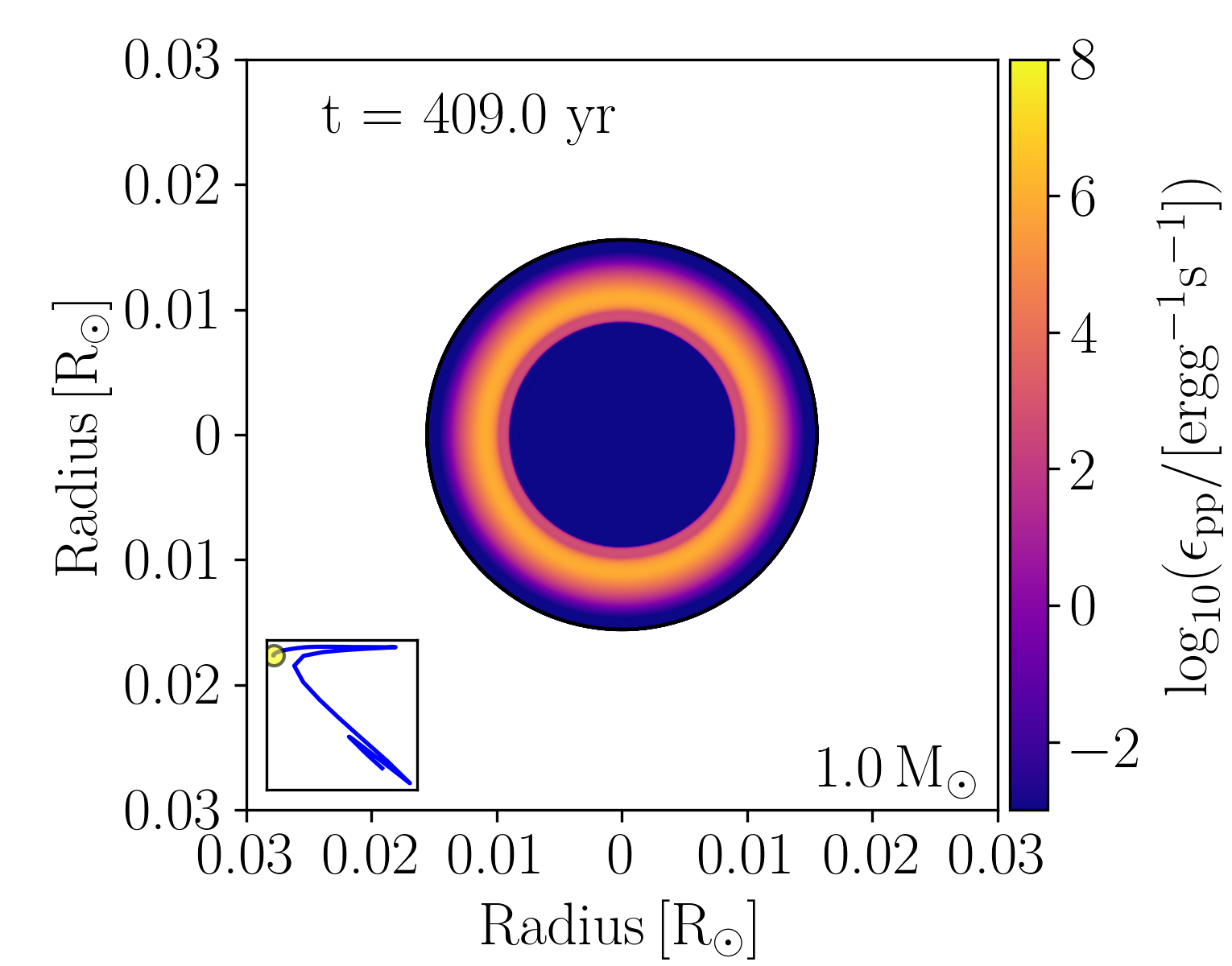}
	\caption{Snapshots of a \texttt{TULIPS} animation that shows the evolution of a 1\Msun white dwarf that is accreting hydrogen-rich material stably at $ 2 \times 10^{-7} \mathrm{M}_{\odot}\,\mathrm{yr}^{-1}$. The diagrams show the radius evolution of the white dwarf. Colors indicate the energy generated by hydrogen burning through the proton-proton chain. From steps 2-4 the white dwarf experiences a violent nova event which leads to an increase in radius of an order of magnitude, exceeding the plot limits.}
	\label{fig:wd_accretion}
\end{figure*}
TULIPS can help understand the consequences of physical processes, such as accretion, on stellar structures. Fig. \ref{fig:wd_accretion} shows how the stellar structure of a 1~\Msun CO white dwarf reacts to accretion of hydrogen-rich material at a rate of $ 2 \times 10^{-7} \mathrm{M}_{\odot}\,\mathrm{yr}^{-1}$ (\texttt{wd\_stable\_h\_burn} model in the MESA test suite). The figure contains snapshots of a property profile animation that visualizes both the evolution of the radius of the white dwarf and of hydrogen burning in its interior (the animation can be found in the supplementary material). The radial direction is proportional to the radius of the white dwarf and the color indicates the values of the specific energy generation rate due to hydrogen burning from the proton-proton chain reaction, which is the dominant burning process. Step 1 shows the initial model of the white dwarf. The dark purple color indicates that no hydrogen burning is taking place initially. In step 2, hydrogen-shell burning is triggered in the white dwarf due to the accretion of hydrogen-rich material. In step 3, the white dwarf experiences a nova: the hydrogen burning intensifies and the released heat causes the radius to increase. By step 4, the radius has increased by more than an order of magnitude, and the outer radius is so large that it exceeds the edges of the plotted region. At this moment, the specific energy generation rate of hydrogen-burning decreases drastically. In step 5, the radius of the white dwarf decreases after the nova and the hydrogen-burning luminosity increases again. From step 5 to the last step, the white dwarf experiences stable hydrogen shell-burning. The radius of the white dwarf, while larger than in the beginning of the simulation, remains constant.

\subsection{Animations to understand the chemical evolution of stellar objects: low metallicity}
\label{sec:animations:lowZ}
\begin{figure*}[ht] 
	\includegraphics[width=0.33\textwidth]{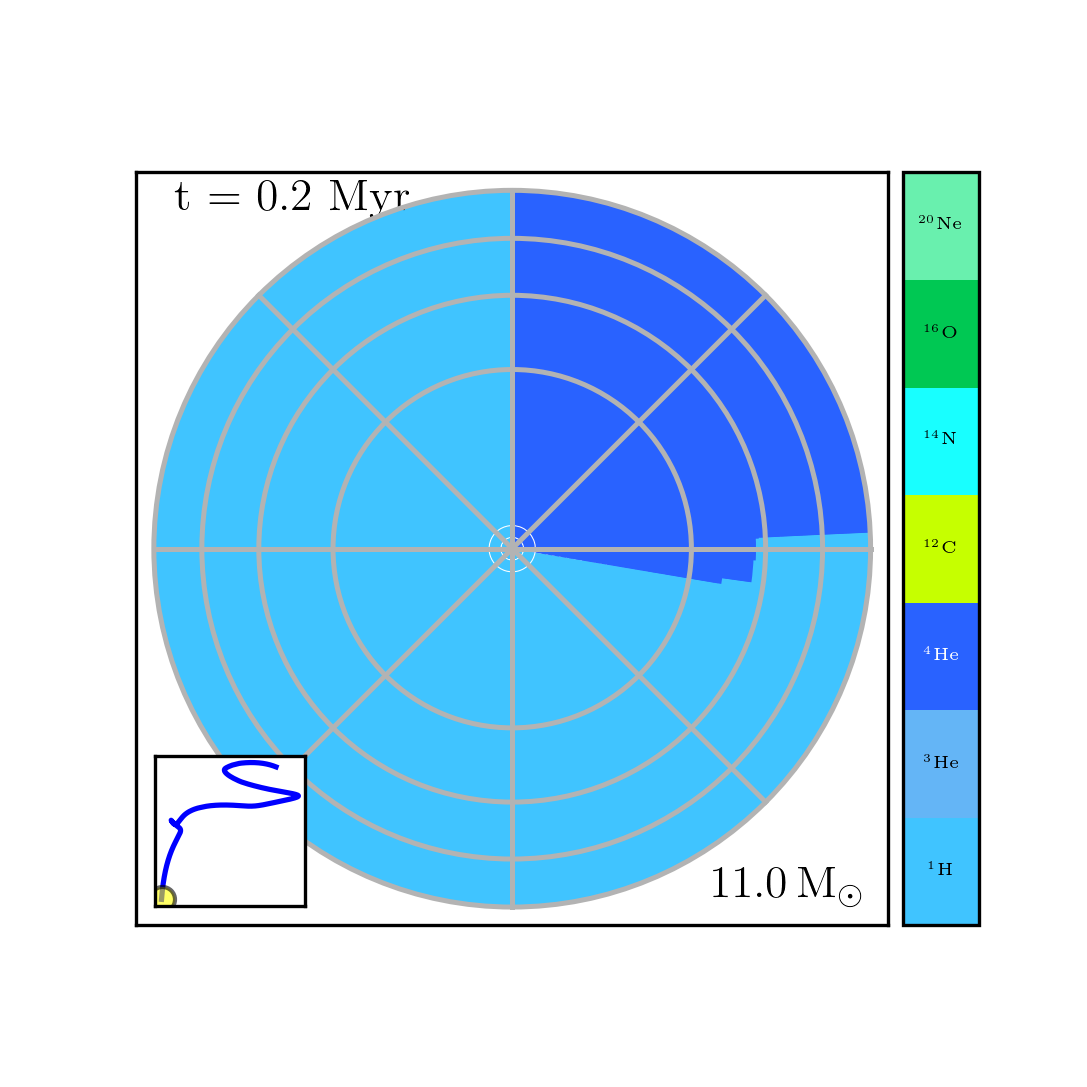}%
	\includegraphics[width=0.33\textwidth]{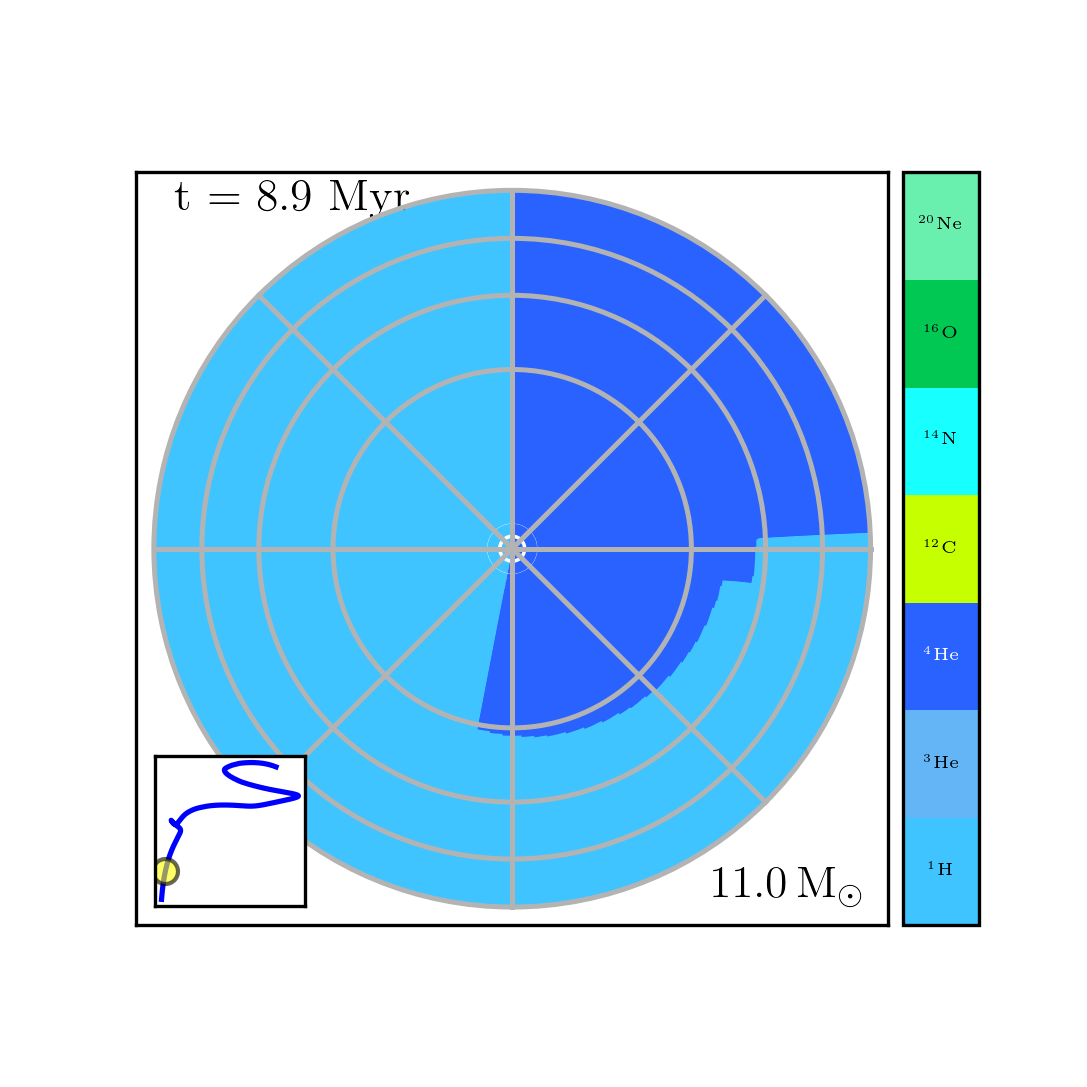}%
	\includegraphics[width=0.33\textwidth]{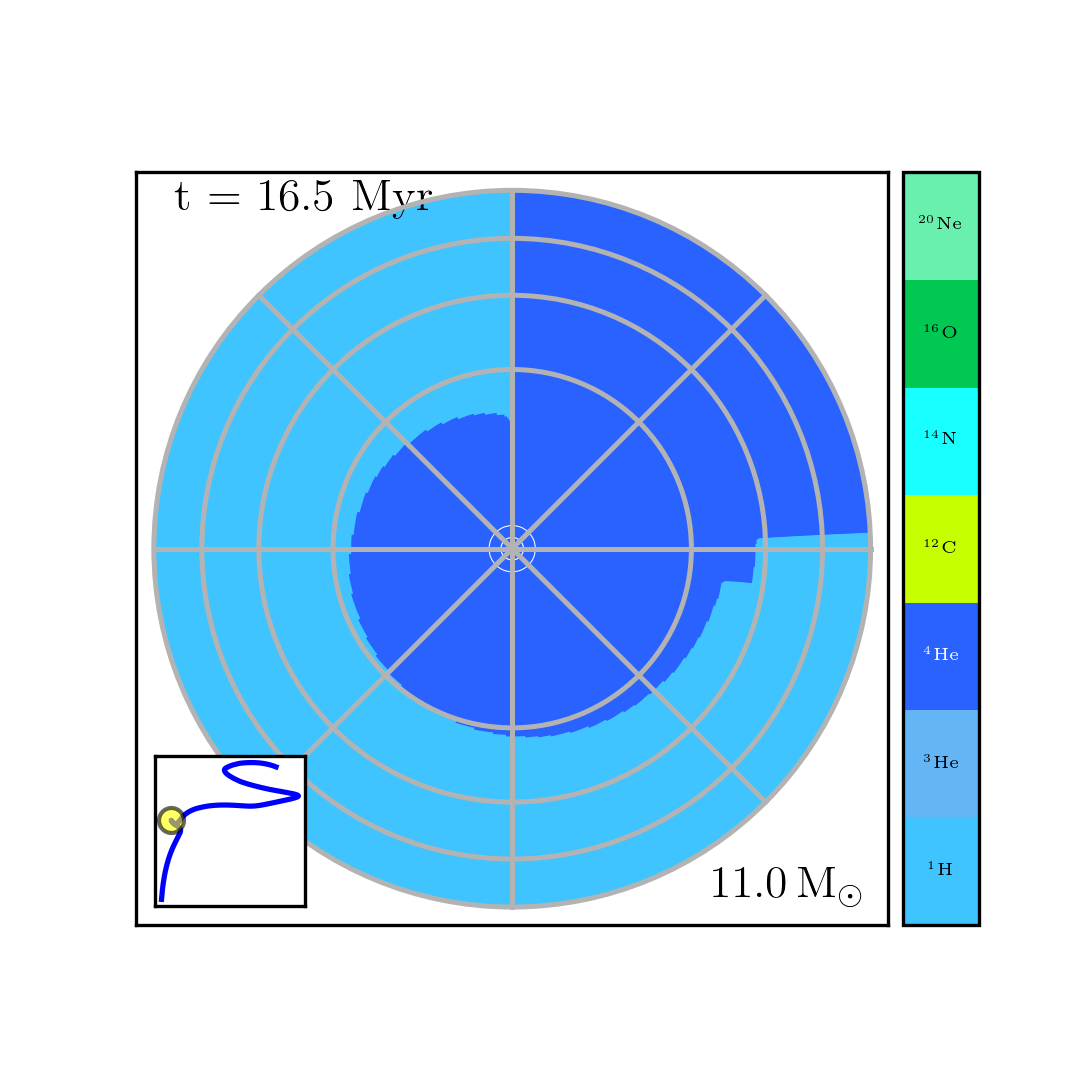}
	\includegraphics[width=0.33\textwidth]{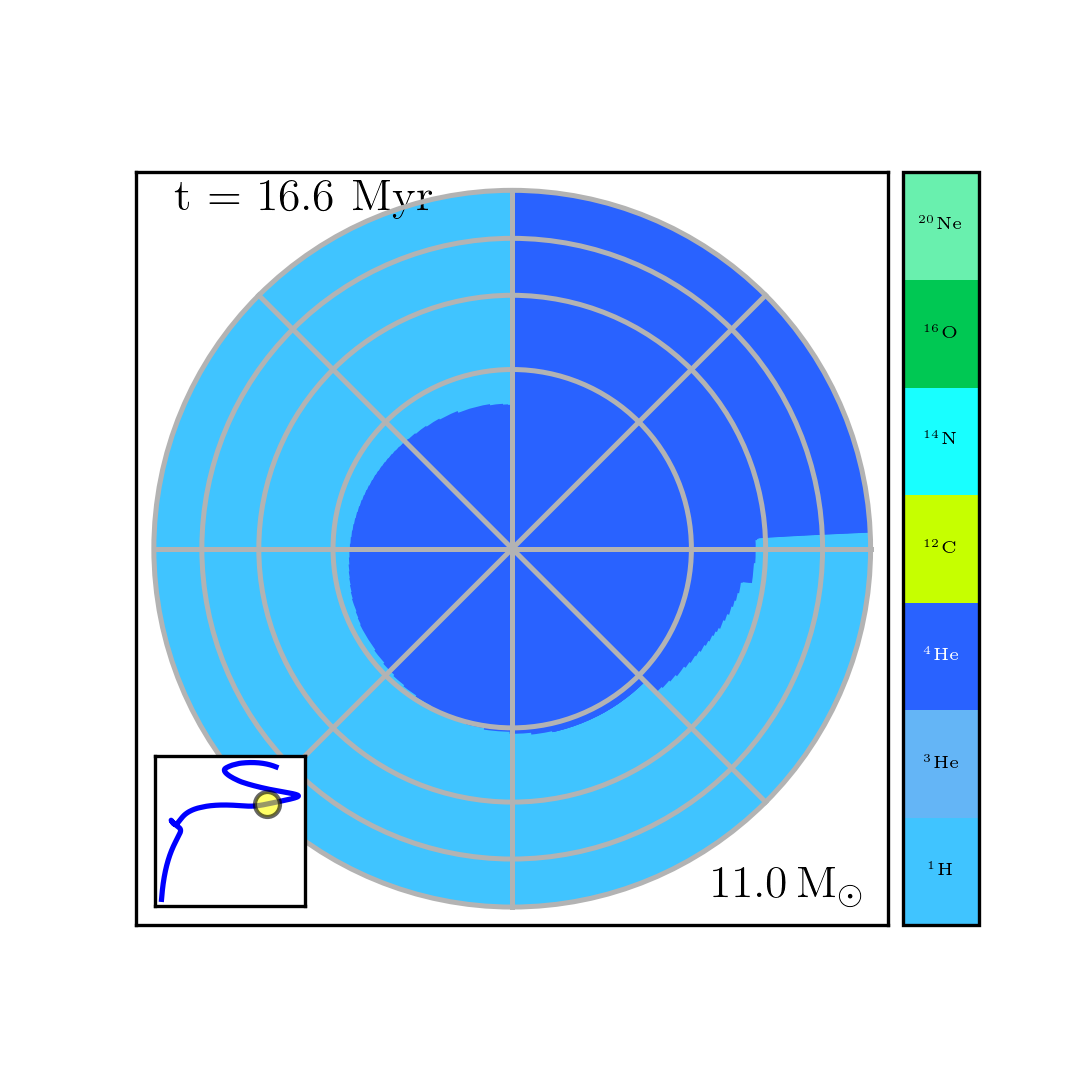}%
	\includegraphics[width=0.33\textwidth]{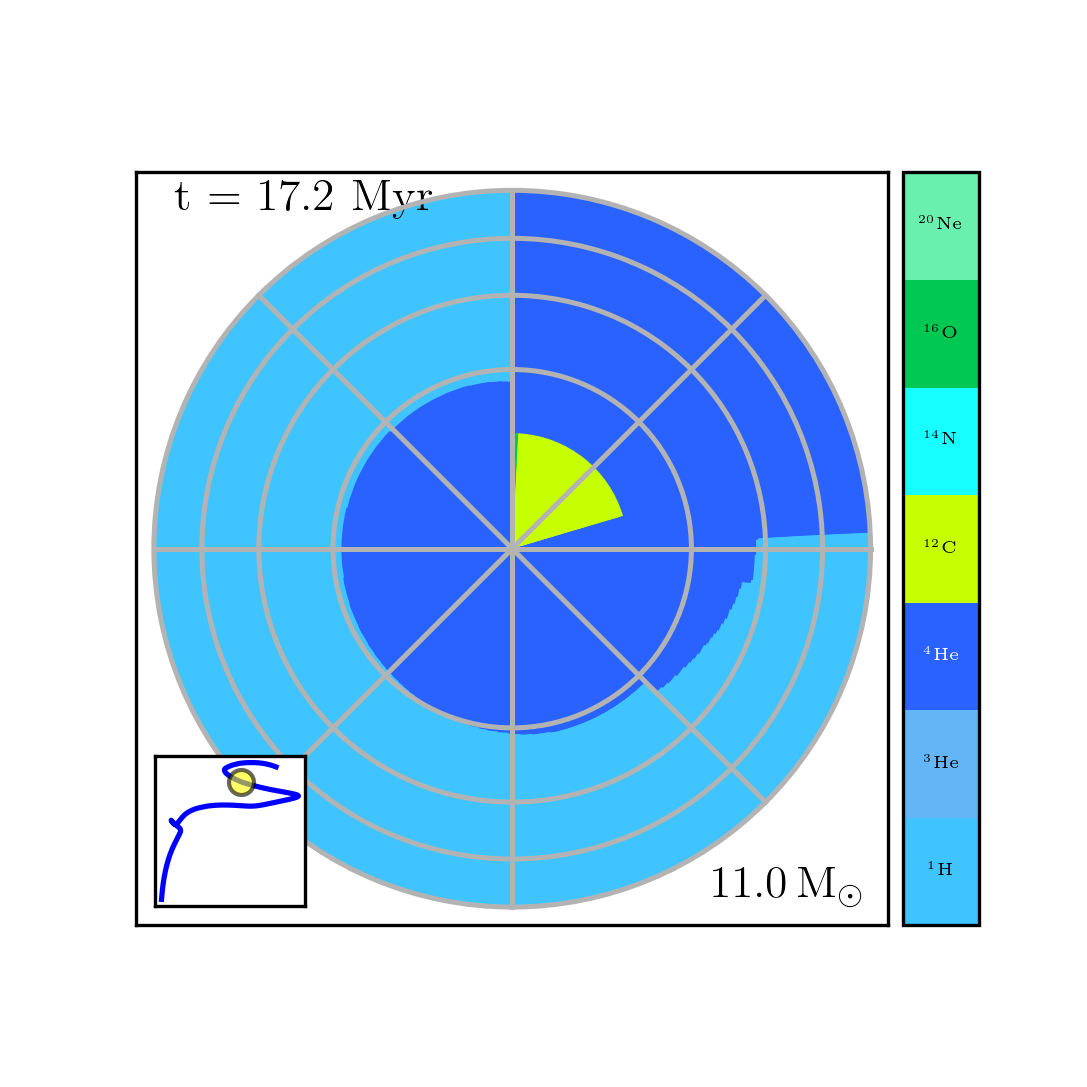}%
	\includegraphics[width=0.33\textwidth]{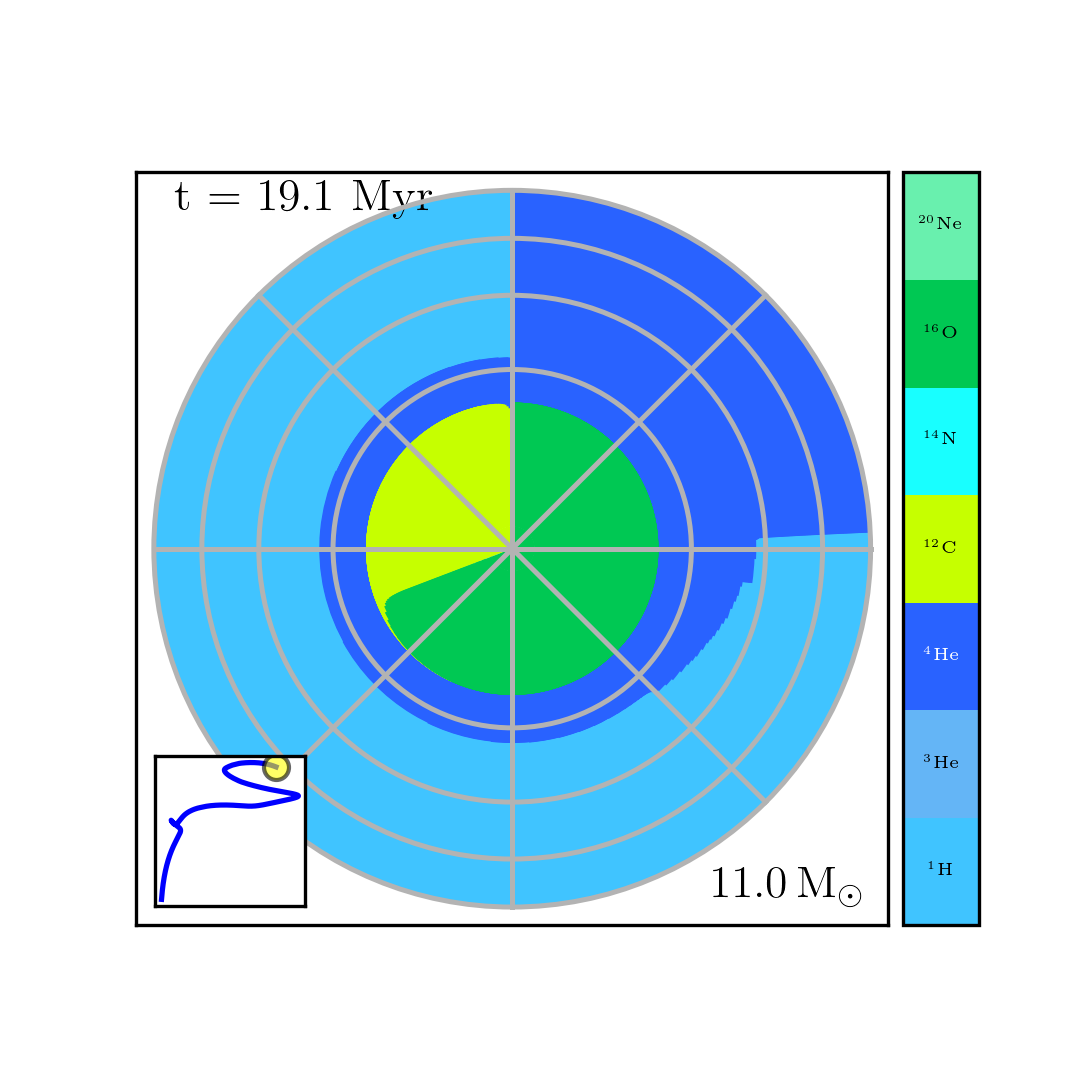}
	\caption{Snapshots of a \texttt{TULIPS} animation that shows the evolution of an 11\Msun star at low metallicity (Z=0.001) until the end of core helium burning. The diagrams show how the composition of the star changes over time. Gray circles indicate the location of mass coordinates at 0.25, 0.5, 0.75, and 1 times the total mass of the star, respectively.}
	\label{fig:lowZ_ev}
\end{figure*}

A major goal of stellar astrophysics is to understand the chemical evolution of the Universe. Stars are the main drivers of this chemical evolution, as they create new elements in their interior while they evolve and eject them into their surroundings through winds and outflows \citep[][]{burbidge_synthesis_1957,woosley_evolution_2002,heger_how_2003,hopkins_galaxies_2014}. With \texttt{TULIPS}, the interior evolution of stars can be visualized in a simple manner. This is illustrated in Fig. \ref{fig:lowZ_ev}, which shows snapshots of a \texttt{chemical\_profile} animation of the evolution of an 11\Msun star at low metallicity (Z=0.001) until the end of core helium burning (the animation can be found in the supplementary material).
The first snapshot shows the onset of hydrogen burning in the star. The outermost layers contain initial hydrogen mass fraction of 0.76 and a helium mass fraction of 0.24. The mass fractions of heavier elements are so small that they are barely visible in the diagram since they are located behind the vertical line. In the region that contains about half the total mass of the star (indicated by the second gray circle from the center), the mass fraction of helium is slightly larger than a quarter, indicating that hydrogen burning has just begun. The second snapshot shows the effect of core hydrogen burning on the composition of the star. After more than 8 Myr, more than half of the hydrogen has been fused into helium nuclei in the stellar core. The mass fraction of hydrogen is decreasing in favor of the helium mass fraction, creating a clockwise motion in the animation. The mass extent of the convective hydrogen-burning core is decreasing due to changes in the opacity, which leads to a smaller extent of the helium-rich region at the bottom of the diagram (6 o'clock) compared to the right-hand side (4 o'clock). In the next snapshot (step 3 in Fig. \ref{fig:lowZ_ev}), the composition profile at the end of the main sequence is shown. Almost the entire core region is now composed of helium. The spiral form of the helium-rich region in the center of the diagram is a consequence of the shrinking of the hydrogen-burning core during the main sequence. Step 4 in Fig. \ref{fig:lowZ_ev} shows the next evolutionary phase, hydrogen-shell burning, which only has a very modest effect on the stellar composition profile: the helium mass fraction outside the helium-rich core increases slightly. In the next snapshot (step 5 in Fig. \ref{fig:lowZ_ev}), carbon and oxygen mass appear in the center, marking the start of core helium burning. In the animation, a clockwise motion can be observed in the center once again. Carbon is created first and in larger amounts than oxygen through triple-alpha captures. When the mass fraction of helium is only about 10\%, the mass fraction of oxygen starts to overtake that of carbon. At the same time, the dark blue region becomes larger, reflecting the mass increase of the helium core up to about a quarter of the total stellar mass. In the last snapshot, the star has reached the end of core helium burning and its core mass is comprised of about 11/16 oxygen and 5/16 carbon. The outer edge of the helium core displays a smooth gradient in composition due to the effect of convective mixing in the envelope.

\section{Comparison between \texttt{TULIPS} and classic diagrams}
\label{sec:comparison}
\texttt{TULIPS} diagrams provide an alternative to classic representation of the physical properties of stars. Here, we compare diagrams created with TULIPS with classic representations of the same properties.

\subsection{Composition of a star}
\label{sec:comparison:composition}
\begin{figure*}[h] 
	\includegraphics[]{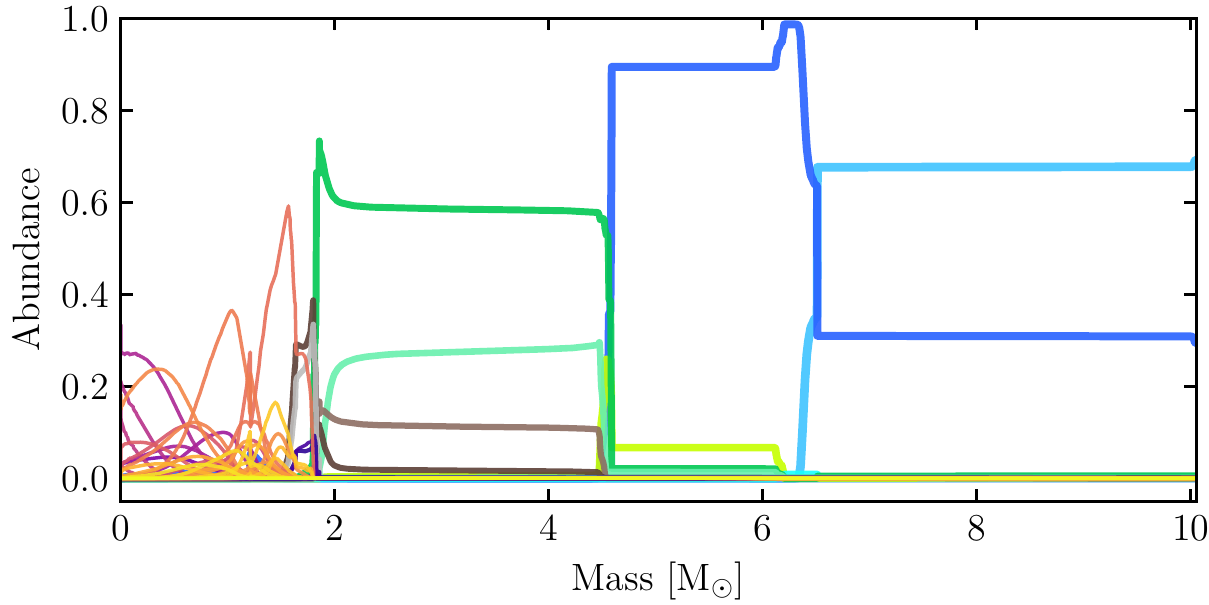}
	\includegraphics[]{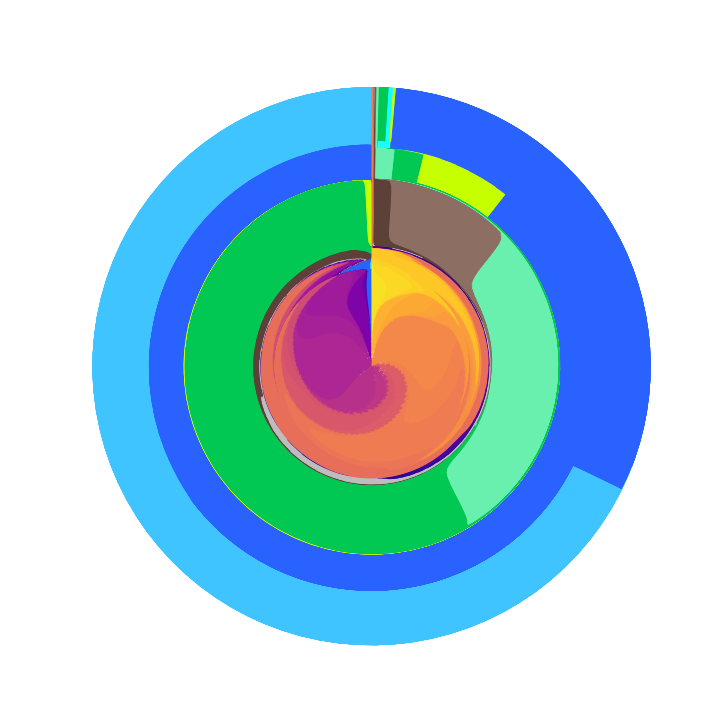}
	\includegraphics[]{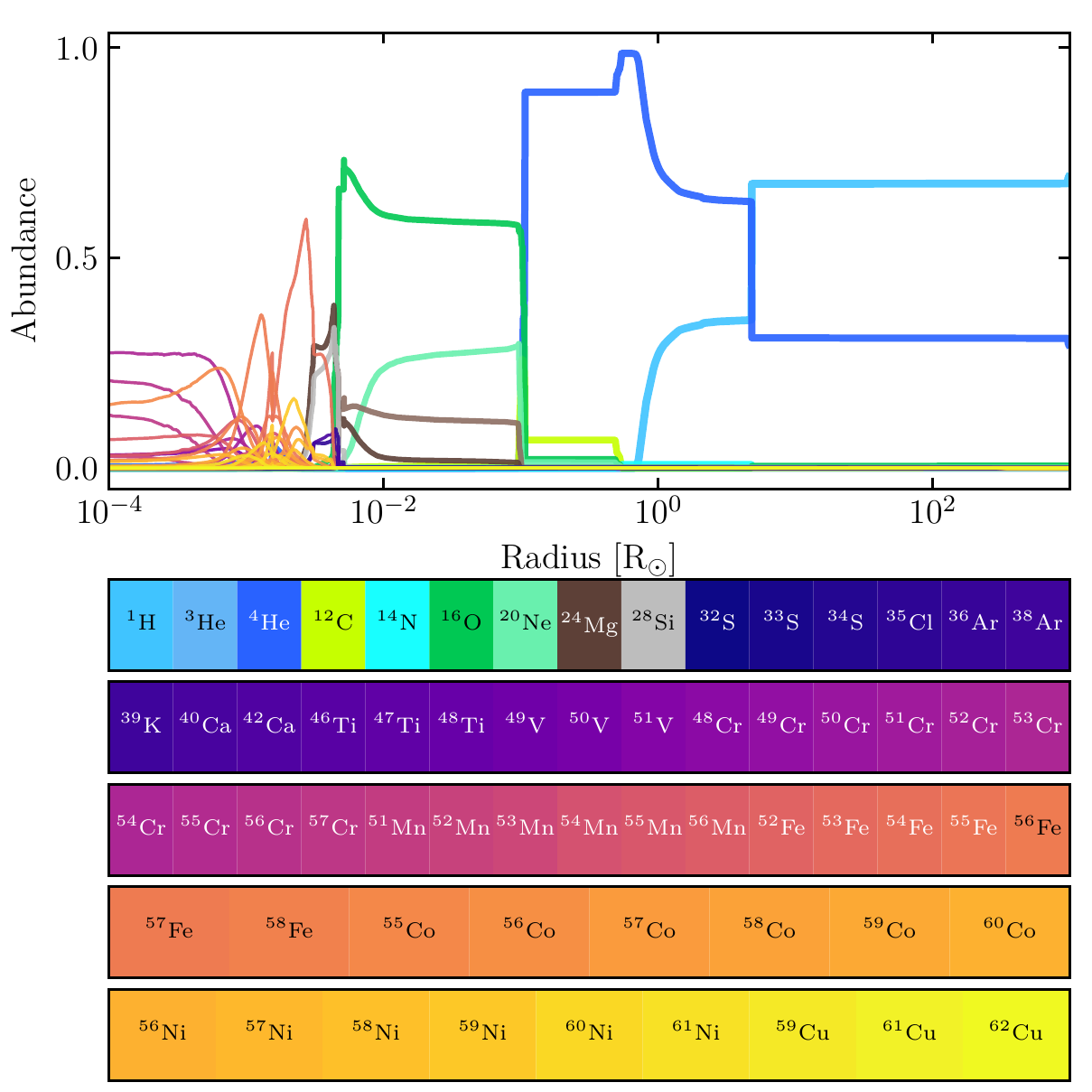}
	\includegraphics[]{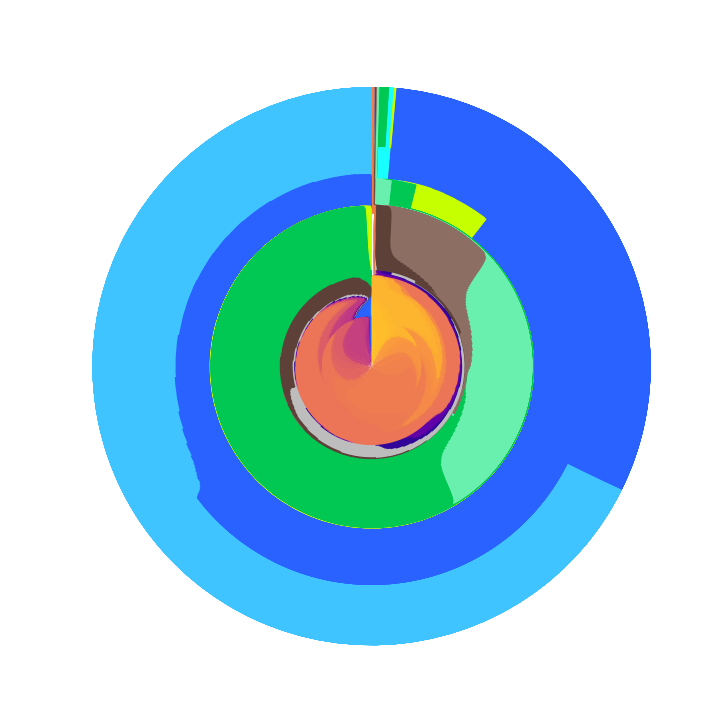}
	\caption{Comparison between classical stellar interior composition diagrams (left) and  diagrams produced with \texttt{TULIPS} (right), containing the same information. All diagrams show the composition of a single star model at solar metallicity with an initial mass of 16\Msun at the onset of core collapse. For clarity, out of the 128 isotopes, only those with maximum mass fractions greater than $10^{-4}$ are shown. The upper panels give the stellar composition as a function of the mass coordinate, and the lower panel as a function of the radial coordinate (on a logarithmic scale) for both the \texttt{TULIPS} and classical diagrams.}
	\label{fig:comparison_composition}
\end{figure*}
In Fig.~\ref{fig:comparison_composition}, classic composition diagrams are compared to corresponding TULIPS diagrams, based on both the radius and the mass coordinate. The difference between the two visualizations is immediately apparent. The classic representation requires more space to be readable and contains a large number of cluttered lines. It facilitates the reading of quantitative information, in particular for individual isotopes. However, reading this information is difficult in the cluttered region in the innermost solar mass that contain a large number of lines representing various isotopes. For the untrained eye, finding distinct layers in the stellar interior requires effort. 

The \texttt{TULIPS} representations in Fig.~\ref{fig:comparison_composition} enable a simple overview of the distribution of isotopes in the stellar interior. Four distinct layers can immediately be identified in the form of distinct rings and circles containing from outside, moving inward, a hydrogen-rich layer, a helium-rich layer, an oxygen/neon layer, and an inner iron-rich core. The mass fraction of every isotope can be read off easily for each ring. For example, in the outermost layers, the mass of the star is divided as 0.75 hydrogen, and approximately 0.25 of helium, which corresponds to a solar-like composition \citep{asplund_chemical_2009} and approximately to most zero-age main-sequence stars in the Universe. In contrast to the classic representation, all isotopes are shown and can be identified, even in the innermost layers. However, reading the exact mass fraction of an element at a particular mass or radius coordinate is more challenging in these diagrams. The rings help give general information about the mass or radius, but it is not possible to find the mass coordinate of a particular composition boundary.

\subsection{Comparing the interior structure of stars}
\label{sec:comparison:interior_prop}
\begin{figure*}[ht!] 
	\includegraphics[]{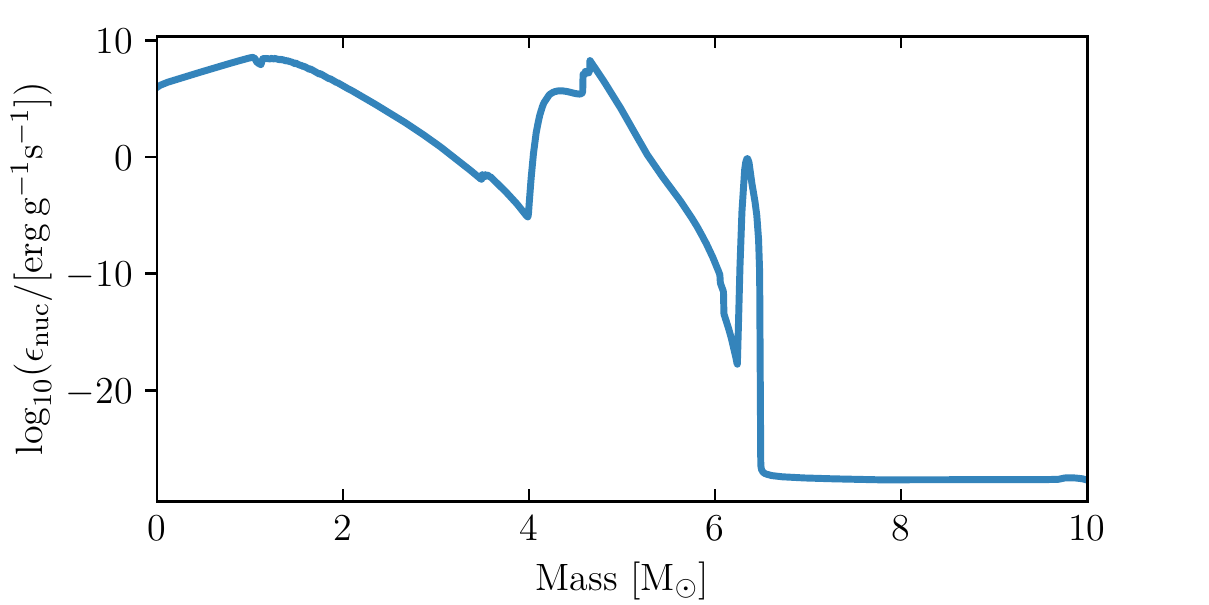}\hspace{-0.9cm}
	\includegraphics[]{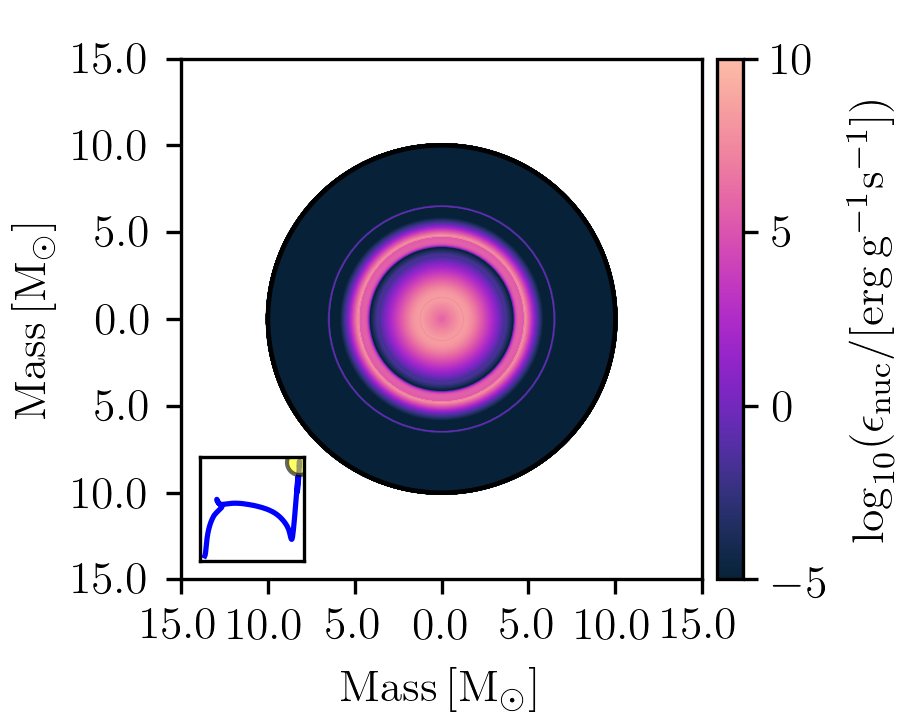}
	\caption{Comparison between a typical diagram showing the nuclear energy generation rate profile of an 11\Msun star at the end of core carbon burning (left) and a diagram produced with \texttt{TULIPS} (right) that contains the same information.}
	\label{fig:comparison_prop}
\end{figure*}

In Fig.~\ref{fig:comparison_prop} we compare a typical representation of two interior properties of a 16\Msun single stars with \texttt{TULIPS} diagrams of the same properties. Both diagrams present the specific nuclear energy generation rate as a function of the mass coordinate. In the standard representation, the quantity is visualized by a line. This figures enables a fast reading of quantitative information, but connecting the variations in the property shown to the overall stellar structure (e.g., realizing that the peak in nuclear energy generation rate corresponds to a shell-burning structure in the star) can be challenging for the untrained eye.

Instead of lines, \texttt{TULIPS} represents physical properties as a color gradient on a circle. This diagram emphasizes the mass coordinates for which the properties shown reach their highest values. For example, in the case of the nuclear energy generation rate, three regions can easily be identified. With these diagrams, it is immediately apparent that the energy generation is located in shells inside the star. Reading quantitative information on the mass coordinate is facilitated by the central and outside axes. However, with this representation, it is generally more challenging to make out quantitative information about the property itself.

\subsection{Interior energy generation and losses, and mixing processes: comparison  with the Kippenhahn diagram}
\label{sec:comparison:kipp}
\begin{figure*}[!ht] 
	\includegraphics[width=\textwidth]{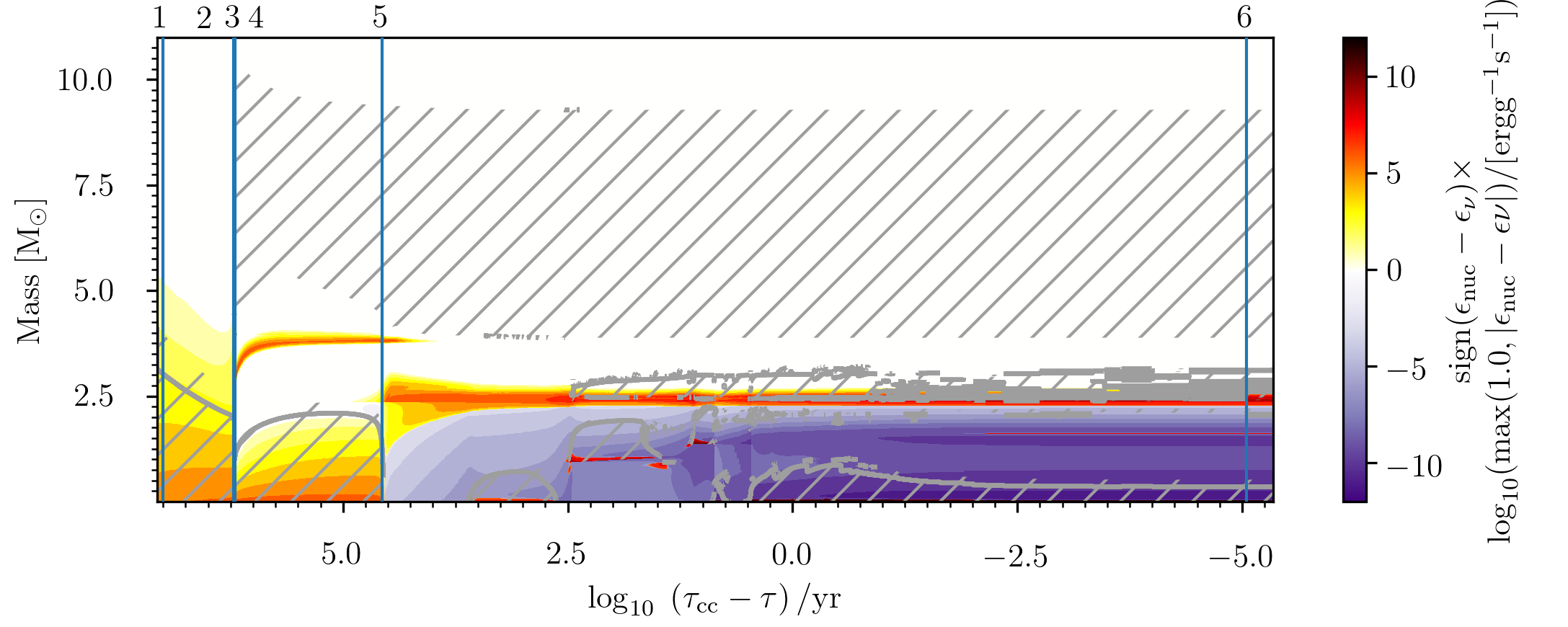}
	\includegraphics[width=0.33\textwidth]{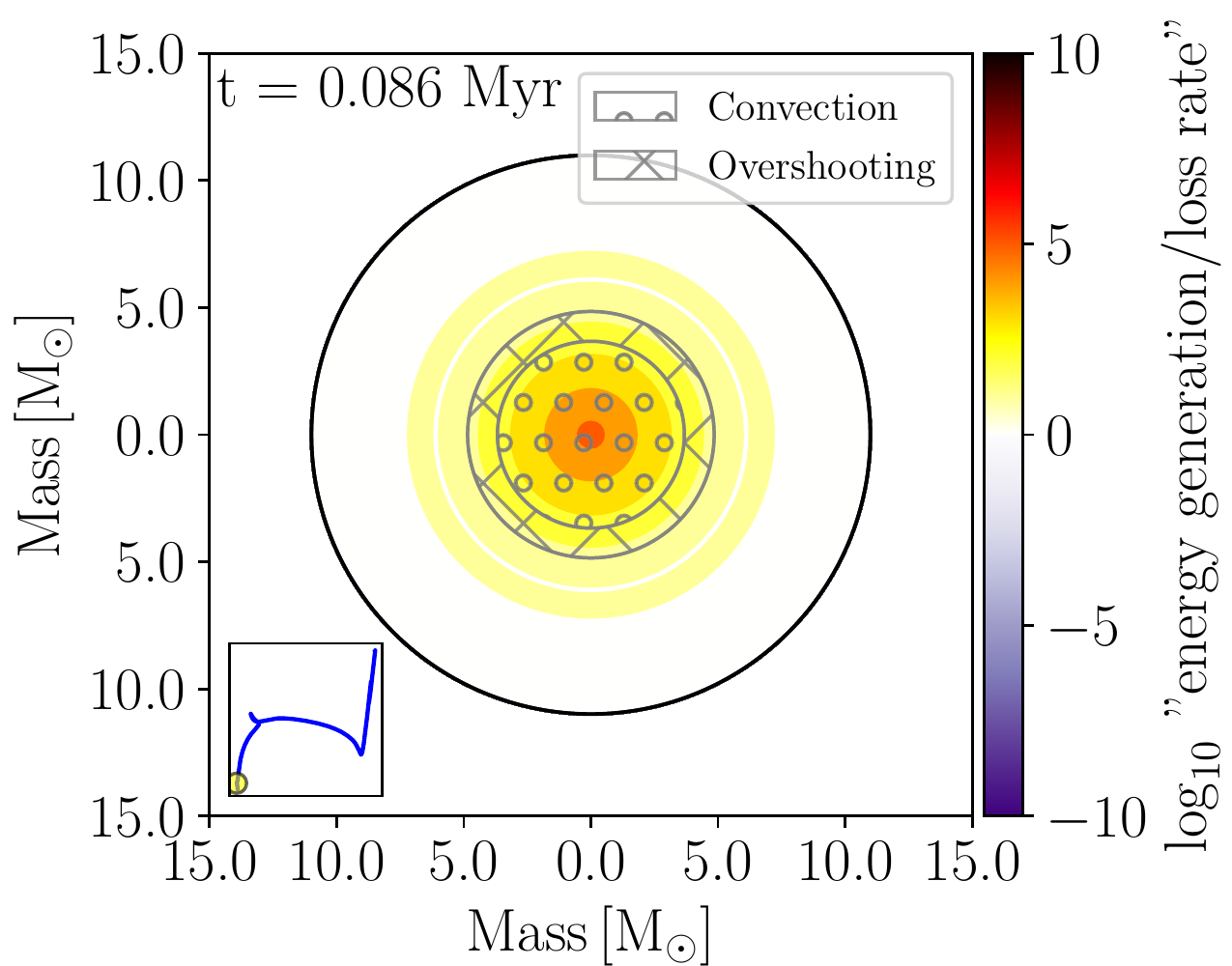}%
	\includegraphics[width=0.33\textwidth]{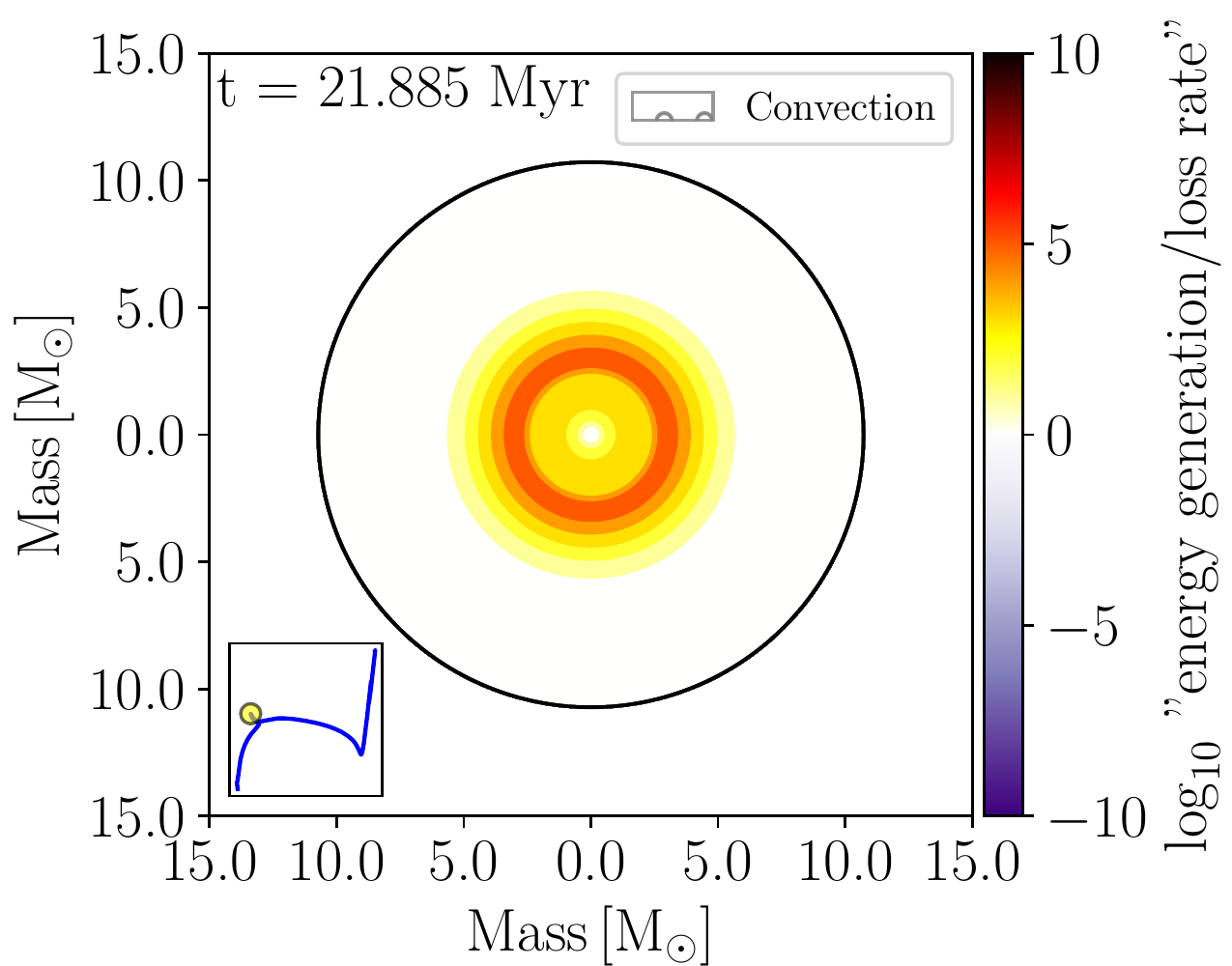}%
	\includegraphics[width=0.33\textwidth]{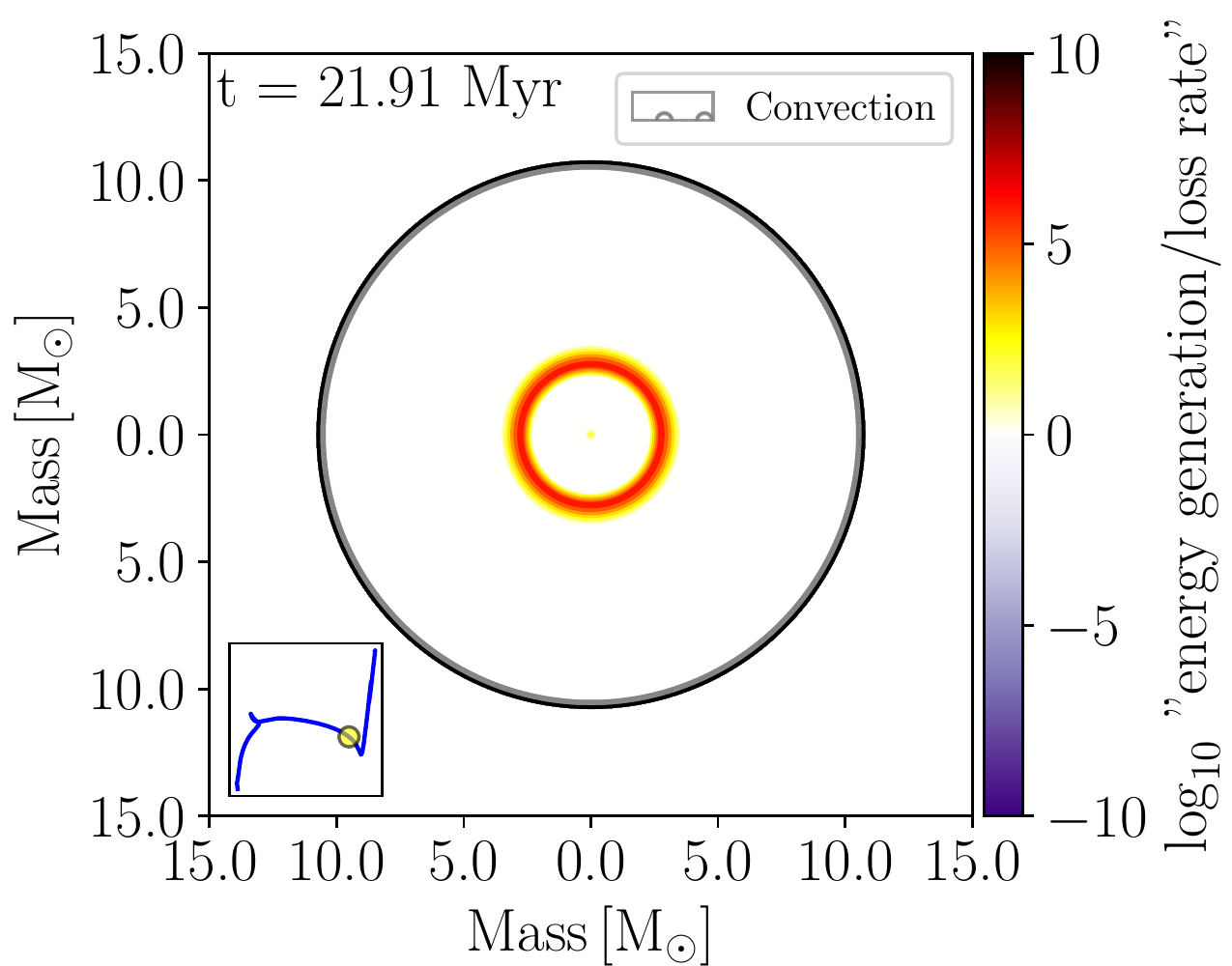}
	\includegraphics[width=0.33\textwidth]{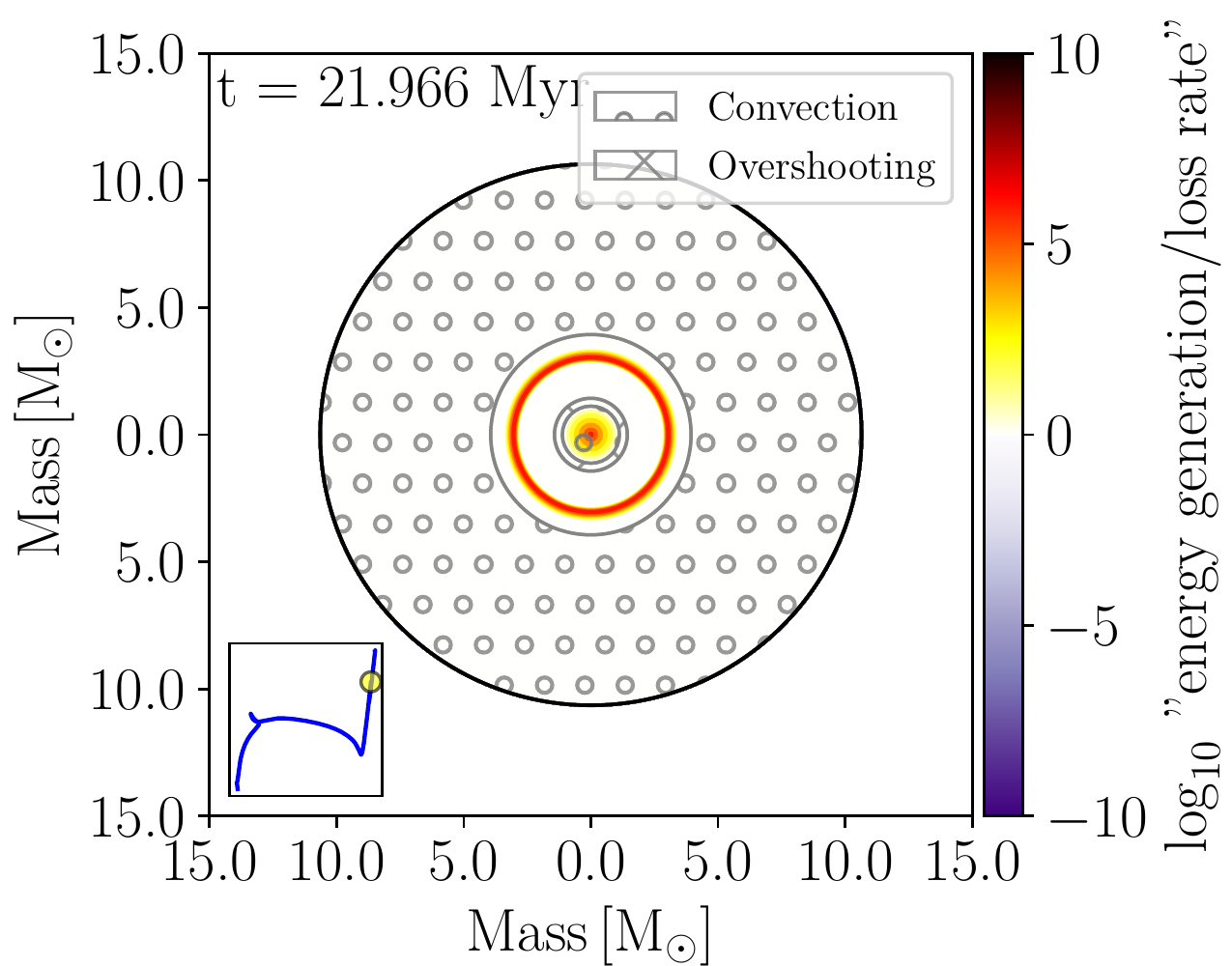}%
	\includegraphics[width=0.33\textwidth]{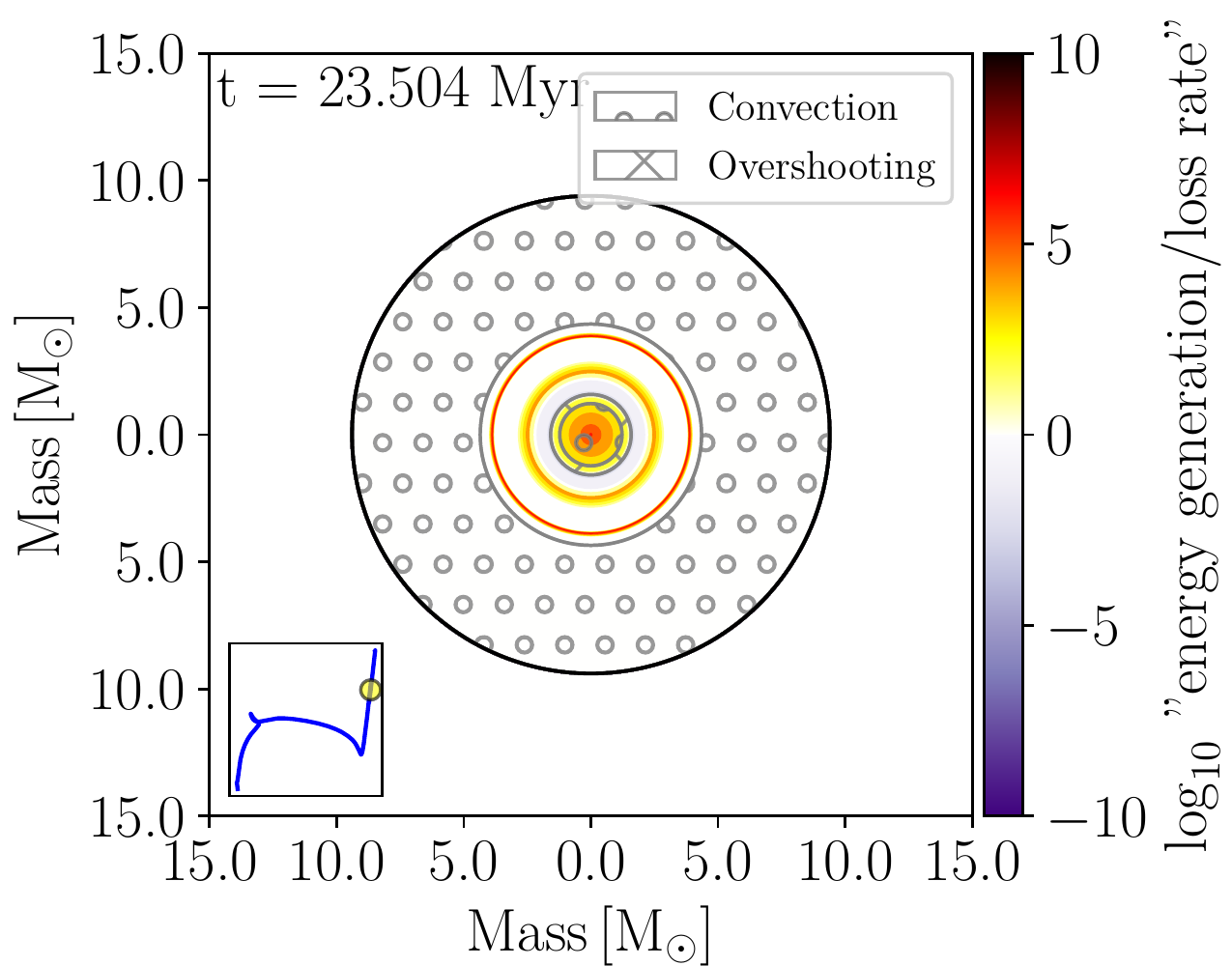}%
	\includegraphics[width=0.33\textwidth]{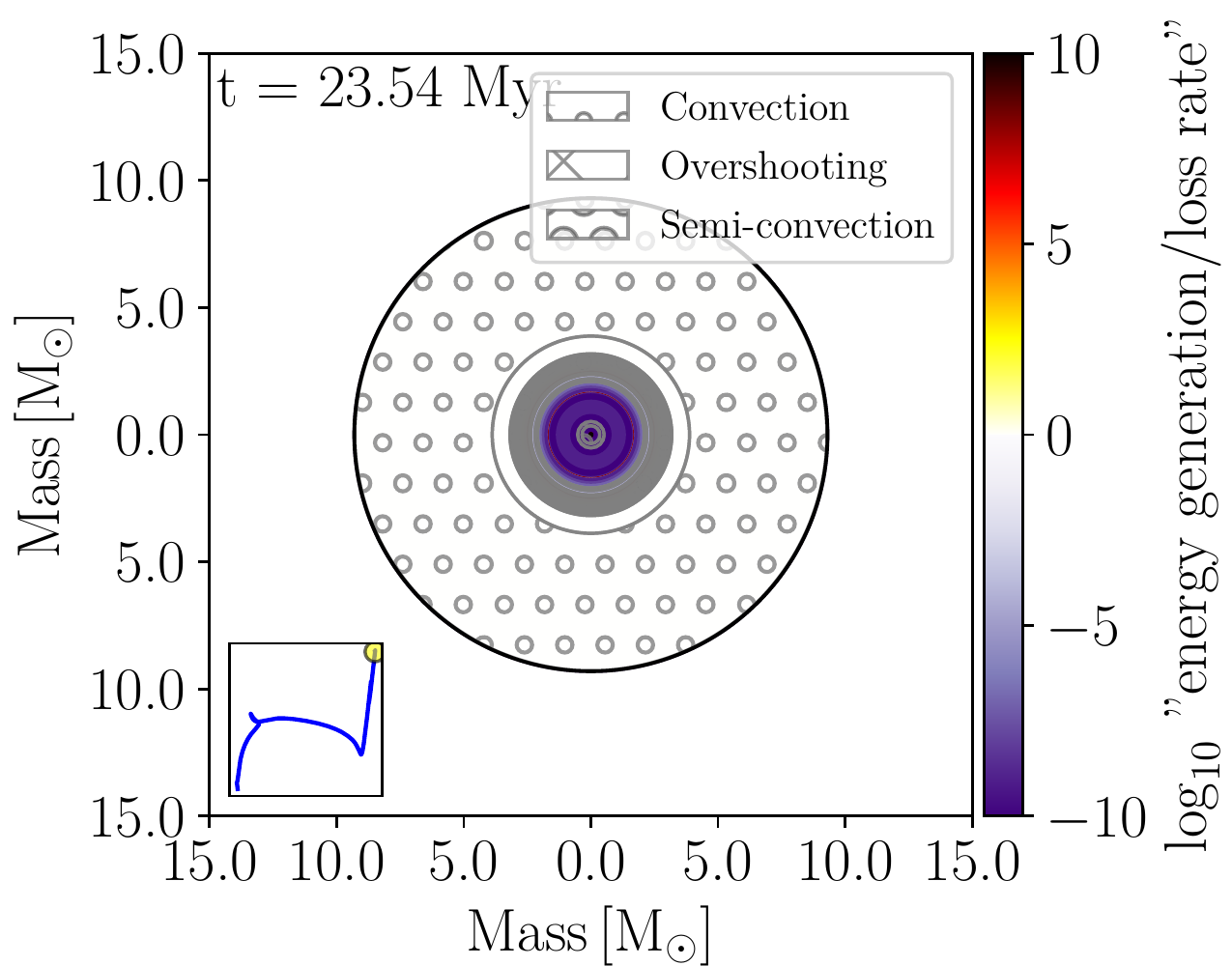}%
	\caption{Comparison of a Kippenhahn diagram (top) to snapshots of a \texttt{TULIPS} animation that shows the energy generation and losses, together with mixing regions in the stellar interior (bottom). These plots showcase the evolution of the same single 11\Msun star model at solar metallicity. In the TULIPS snapshots, the stellar model is represented by a circle whose radius represents the total mass of the model. Inset diagrams indicate the evolution on the HRD. In both the Kippenhahn diagram and the TULIPS diagrams, convective regions are indicated by hatching and colors indicate the logarithmic difference between the specific nuclear energy generation rate $\epsilon_{\mathrm{nuc}}$ and the specific energy rate due to neutrino emission $\epsilon_{\mathrm{\nu}}$. Blue vertical lines on the Kippenhahn diagram mark the moments at which the TULIPS snapshots were generated.}
	\label{fig:kipp_m11}
\end{figure*}

\texttt{TULIPS} animations can be used to better interpret the evolution of the energy generation and losses, as well as mixing processes in the interior of stellar objects. This evolution is traditionally shown with Kippenhahn diagrams \citep{hofmeister_sternentwicklung_1964} that represent the interior stellar structure as vertical profiles as a function of time, where the vertical direction is proportional to the mass coordinate. In Fig.~\ref{fig:kipp_m11}, we compare the Kippenhahn diagram of our example model of an 11\Msun star (see also Section \ref{sec:animations:HRD}) to \texttt{TULIPS} animations that show the interior burning and mixing processes particular moments in time that are marked by vertical lines on the Kippenhahn diagram.

The Kippenhahn diagram is information-dense and allows an overview of the burning and mixing processes throughout the evolution. Distinct core burning stages can be easily distinguished by the advent of a convective region in the center of the diagram, where energy is being generated. Short phases are difficult to distinguish on this diagram, as are details of the burning and mixing structure at one given moment in time.
The snapshots of the corresponding \texttt{TULIPS} animation represent the star as a circle that is divided into rings. Colors and hatching indicate the burning and mixing regions, respectively. In the first snapshot, the star contains a convective hydrogen-burning core. The animation visualizes how the convective regions change in mass extent over time and shows that shell burning regions occupy a small fraction of the total mass of the star. The change in mass extent of convective regions can be easily recognized by following the evolution of regions that are shaded with small circles. After core carbon depletion (snapshot 5), energy losses due to neutrino emission clearly dominate in the core. The \texttt{TULIPS} representation also highlight the complex variations in energy generation and losses towards the end of the evolution of such a massive star. However, it does not enable a quick overview of all burning process that occur in the star over its lifetime.

\section{Discussion, conclusion, and outlook}
\label{sec:disc_conclusion}
\texttt{TULIPS} is a novel visualization tool for stellar astrophysics. It enables intuitive visualizations of the physical properties of stellar objects based on state-of-the-art one-dimensional stellar evolution simulations. It is currently optimized for stellar models computed with the MESA code. Making use of the intrinsic assumption of spherical symmetry in one-dimensional simulations, \texttt{TULIPS} represents the physical properties of stellar objects as the properties of circles. The basic functionalities of \texttt{TULIPS} include creating diagrams that visualize (a) the size and apparent color of stellar objects, (b) the interior burning and mixing processes, (c) the composition of stellar objects, and (d) the interior physical properties. The heart of \texttt{TULIPS}' capabilities consists in creating interactive animations that show the time evolution of these diagrams and the physical properties they represent.

This paper demonstrates how \texttt{TULIPS} can be used as an analysis tool for understanding the evolution of stellar objects. It shows that \texttt{TULIPS} can help to visualize physical processes, such as accretion and mixing processes. 
Compared to classic representations, \texttt{TULIPS} diagrams typically require less space, easily represent qualitative information, and help appreciate the scales of stellar objects. These diagrams are better in line with design principles for data visualization \citep{evergreen_design_2013}, closer to the actual shape of these objects, and as such, probably more intuitive \citep{lewis_culture_2006}. However, quantitative information is harder to apprehend with \texttt{TULIPS} diagrams. The amount of information that can be conveyed is limited and they do not allow an overview of the global time evolution of physical quantities. Research in data visualization indicates that while dynamic visualizations are better at drawing and holding attention, making it more evocative, deeper understanding can be gained with static ones \citep{valkanova_reveal-it_2013,newell_picture_2016}.

In summary, \texttt{TULIPS} does not replace the classic representations of physical properties, but allows complementary insight and a change of perspective. This in turn has the potential to trigger and hold the attention of the readers \citep{treisman_features_1988,hillstrom_visual_1994}. Because they can convey the same information in a simpler manner and attention has been given to increasing their appeal according to design principles, these diagrams hold the potential to improve communication and understanding of stellar astrophysics \citep{lusk_effect_1979,cawthon_effect_2007,evergreen_design_2013}. 

Aside from a tool for research, \texttt{TULIPS} can be applied as a means to teach others the evolution and structure of stellar objects and also as a means to produce material for public outreach. 

In the future, \texttt{TULIPS} could be further extended, for example by including the evolution of binary stars, by enabling more interactions with the diagrams, and by adapting it for use with other one-dimensional physical simulations.

\section*{Acknowledgments}
The author thanks the anonymous reviewers for helpful suggestions that improved the manuscript. The author is grateful to S. Justham for important insights and suggestions that led to improvements of the project, and for devising the unforgettable acronym. This work has benefited from valuable input and guidance by S. E. de Mink. The author thanks R. Farmer for making the \texttt{mesaPlot} package open source, and for help with debugging and testing of the code. L. Kaper and A. de Koter provided helpful feedback on the manuscript. The author thanks the entire BinCosmos/BinWaves group for helpful suggestions and testing. This project was funded by the European Union’s Horizon 2020 research and innovation program from the European Research Council (ERC, grant agreement No.715063) and the Netherlands Organisation for Scientific Research (NWO) as part of the Vidi research program BinWaves with project number 639.042.728. The author gratefully acknowledges the ET Outreach award of the Royal Holland Society of Sciences and Humanities and its sponsors E. van Dishoek and T. de Zeeuw that made it possible to fund the excellent work of I. de Langen on the TULIPS documentation and tutorials. The author is grateful to A. Faber for designing a logo that captures the essence of TULIPS. Finally, the author thanks B. Sutlieff, L. A. C. van Son, and H. J. G. L. M. Lamers for their unwavering enthusiasm and support, and for helpful discussions about TULIPS, without which this project would not have come into bloom. 
\bibliography{tulips.bib,additional.bib}

\end{document}